\documentclass[12pt]{article}
\usepackage{ulem}
\usepackage{latexsym}
\usepackage[dvipdfmx]{graphicx}
\usepackage[top=3cm,bottom=3.5cm,left=2cm,right=2cm]{geometry}
\usepackage{pifont}          
\usepackage{amsmath,amssymb,mathrsfs}
\usepackage{color}
\usepackage{geometry}
\usepackage{caption}
\usepackage{ulem}
\usepackage{setspace}
\usepackage{bm}
\newcommand{\nn}{\nonumber}

\newcommand{\be}{\begin{eqnarray}}
\newcommand{\ee}{\end{eqnarray}}

\newcommand{\h}{\hspace}
\newcommand{\aln}[1]{\begin{align}#1\end{align}}

\begin{document}
\title{
\vspace{-2cm}
\vbox{
\baselineskip 14pt
\hfill \hbox{\normalsize KEK-TH-2208, TU-1105}} 
\vskip 1cm
\bf \Large QCD Axions and CMB Anisotropy
\vskip 0.5cm
}
\author{
 Satoshi Iso$^{ab}$\thanks{E-mail: \tt iso(at)post.kek.jp},\h{1mm} 
 Kiyoharu~Kawana$^{ac}$\thanks{E-mail: \tt kawana(at)post.kek.jp}\h{1mm} and 
 Kengo Shimada$^{d}$\thanks{E-mail: \tt kengo.shimada.c5(at)tohoku.ac.jp}
\bigskip\\
\it 
\normalsize
 $^a$ Theory Center, High Energy Accelerator Research Organization (KEK),\\
 \normalsize 
\it  $^b$ Graduate University for Advanced Studies (SOKENDAI),\\
 \normalsize
\it Tsukuba, Ibaraki 305-0801, Japan, \\
\normalsize 
\it  $^c$ Center for Theoretical Physics, Department of Physics and Astronomy,\\
 \normalsize
\it  Seoul National University, Seoul 08826, Korea,
\\
\normalsize
\it  $^d$ Department of Physics, Tohoku University, Sendai, Miyagi 980-8578, Japan
\smallskip
}
\date{\today}

\maketitle

\abstract{
In this paper, we 
consider a possibility that the temperature anisotropy of cosmic microwave background (CMB) is dominantly generated
by the primordial  fluctuations of QCD-axion like particles 
under a circumstance that inflaton's perturbation is  too small to explain the CMB anisotropy. 
Since the axion potential is  generated by acquiring its energy from radiation, 
the primordial fluctuations of the axion field  generated in the inflation era 
are correlated with the CMB anisotropies.
Consequently, 
the observations stringently constrain a model of the axion and the early universe scenario. 
The following conditions must be satisfied: 
(i) sufficient amplitudes of the CMB anisotropy 
(ii) consistency with the axion isocurvature constraint and  (iii) the non-Gaussianity constraint.  
To satisfy these conditions, a large energy fraction $\Omega_A^{}$ of the axion is necessary at the QCD scale when the axion-potential is generated, but simultaneously, it must become tiny at the present era due to the isocurvature constraint. 
 Thus an additional scenario of the early universe, such as low scale thermal inflation,  is inevitable to 
dilute the axions after the QCD scale. 
We investigate such a model and obtain its allowed parameter region. 
}
\newpage

%_____________________Introduction___________________
\section{Introduction}
Thanks to the development of observational cosmology, we have obtained a lot of information to constrain  models of particle physics beyond the standard model (BSM). 
In particular, the observations of the anisotropy of the cosmic microwave background (CMB) are one of the most fascinating since they are generated at a very high energy scale and possibly related to the BSM. 
The current observational data such as Planck 2018 \cite{Aghanim:2018eyx,Akrami:2018odb,Akrami:2019izv} tells us that the temperature fluctuation is almost scale invariant and adiabatic, which favors inflation models based on a  single scalar field. 
In particular, inflation models in which the Standard Model (SM) Higgs plays a role of an inflaton have been attracting much attention due to its simplicity and phenomenological richness 
%\cite{Bezrukov:2007ep}-\cite{Hamada:2014wna}.
\cite{Bezrukov:2007ep,Bezrukov:2008ej,Bezrukov:2008ut,GarciaBellido:2008ab,Bezrukov:2010jz,Bezrukov:2012sa,Hamada:2014iga,Hamada:2014wna}. 
On the other hand, many of BSMs predict multiple light scalar fields, which may generate isocurvature perturbations. 
A common example is the QCD axion 
%\cite{Peccei:1977hh}-\cite{diCortona:2015ldu} 
 \cite{Peccei:1977hh,Wilczek:1977pj,Weinberg:1977ma,Kim:1979if,Shifman:1979if,Zhitnitsky:1980tq,Dine:1981rt,Kim:2008hd,diCortona:2015ldu} 
 whose Peccei-Quinn  symmetry is already broken before the primordial inflation.  
If the scale of inflation is above the QCD scale, axion is massless and its fluctuation grows during inflation. 
As the universe cools down to the QCD scale, the axion  acquires non-zero energy from coherent configurations of gluons through non-perturbative effects, and the density fluctuations of both the axion and the radiation are generated. 
In the ordinary early universe scenario of the QCD axion 
%\cite{Preskill:1982cy}-\cite{Kawasaki:2013ae}, 
 \cite{Preskill:1982cy,Abbott:1982af,Dine:1982ah,Beltran:2006sq,Hertzberg:2008wr,Wantz:2009it,Hikage:2012be,Kawasaki:2013ae}, 
 the resultant CMB fluctuation has large  isocurvature component and its magnitude is stringently constrained by Planck 2018 \cite{Akrami:2018odb}. 

Although CMB observations rule out purely isocurvature  perturbations, we can consider a possibility that such a primordial isocurvature fluctuation is somehow  converted to adiabatic one at the early universe. 
A well known example is the curvaton scenario 
%\cite{Lyth:2001nq}-\cite{Byrnes:2010xd}
\cite{Lyth:2001nq,Enqvist:2001zp,Moroi:2001ct,Lyth:2002my,Gordon:2002gv,Dimopoulos:2003az,Sasaki:2006kq,Chingangbam:2009xi,Enqvist:2009ww,Nakayama:2009ce,Byrnes:2010xd}
 where the conversion occurs by the decay of curvaton into radiation. 
Generally speaking, 
%not only explaining the amplitude of scalar perturbation of CMB, 
such  conversion mechanisms generate  strong non-Gaussianity even if the primordial  fluctuation is Gaussian 
due to the mechanism itself as well as anharmonicity of the axion potential, and
 we can constrain various models  by  non-Gaussianity. 

In this paper, we consider a possibility that the CMB anisotropy is dominantly generated by the 
 primordial perturbation of a ``{\it QCD-axion like particle}". 
Here, by a QCD-axion like particle, we mean a general class of particles which are (nearly) massless during the inflation and obtain its potential at a relatively low energy scale, e.g., QCD scale,  in  the thermal history of the universe. 
Throughout this paper, we simply call such a field {\it axion}  and denote the temperature at which its  potential 
is produced by $T_A^{}$. 
For the QCD axion, $T_A^{}$ is given by $T_{\rm QCD}^{}$.
In order to realize such a scenario, we need the following conditions for the evolution of the axion field.  
First,  a large amount of axions at $T=T_A$ is necessary to suppress non-Gaussianity. 
Second, axions must be largely diluted until present in order to satisfy the isocurvature constraint. 
In addition, to explain the observed CMB anisotropy, we have a relation between the axion abundance at $T_A$ and 
 $H/f_A \theta$, where $H$ is the Hubble constant of the primordial inflation and $\theta$ is the misalignment angle. 
These conditions cannot be simultaneously satisfied in the standard scenario of QCD axions, but if thermal inflation 
%\cite{Lyth:1995ka}-\cite{Baratella:2018pxi} 
\cite{Lyth:1995ka,Gong:2016yyb,Hambye:2018qjv,Baratella:2018pxi} 
occurs after the QCD temperature, we can construct a model to satisfy them. 

This paper is organized as follows. 
In section \ref{sec:preliminary}, we first give a brief introduction to non-linear curvature perturbations for fixing 
notations. Then in section \ref{axion scenario}, we investigate a possibility of the scenario of 
generating CMB anisotropy from axion fluctuations. In particular, we focus on non-Gaussianity and isocurvature
perturbations, and their observational constraints on model parameters. 
Finally in section \ref{sec:model}, we investigate a few models.
The QCD axion in the standard model cannot satisfy the observational constraints. 
On the other hand, if thermal inflation 
occurs after QCD transition, there is an allowed region in the model parameters. 
We also comment on a classically conformal $B$-$L$ model as an explicit model
 in which the electroweak (EW) symmetry 
is supercooled and thermal inflation naturally occurs around the QCD scale. 
In 
\ref{Curvature perturbations and CMB observables}, we briefly summarize  how 
the curvature perturbations are related to the CMB observables. 
In \ref{app:gradual}, we take effects of gradual energy transfer in the calculations of 
density perturbations, which turns out to give small corrections to the instantaneous 
approximations discussed in the body of the paper. 
In \ref{final fluctuation}, we discuss how the curvature perturbations at the QCD transition
are converted to the present curvature perturbations.  
In \ref{app:isocurvatureNG}, we calculate isocurvature non-Gaussianity.  

%%%%%%%%%%%%%%%%%%%%%%%%%%%%%%%%%%%%%%%%%%
%%%%%%%%%%%%%%%%%%%%%%%%%%%%%%%%%%%%%%%%%%
%%%%%%%%%%%%%%%%%%%%%%%%%%%%%%%%%%%%%%%%%%
%_____________________General_______________________
\section{Preliminaries}
\label{sec:preliminary}
We first summarize basic notions of the non-linear generalizations of curvature perturbations \cite{Sasaki:2006kq,Lyth:2004gb}. 
We then summarize observational constraints from the non-Gaussianity and the isocurvature fluctuations in Planck 2018 \cite{Akrami:2018odb,Akrami:2019izv}.  
See 
%\cite{Brandenberger:2003vk}-\cite{Baumann:2009ds} 
\cite{Brandenberger:2003vk,Bartolo:2004if,Weinberg:2008zzc,Baumann:2009ds} 
for more details about the cosmological perturbations.  

%_____________________Curvature perturbations_______________________
\subsection{Non-linear curvature perturbations}
In discussing the large scale metric fluctuations (i.e. {\it separate universe} hypothesis \cite{Sasaki:1998ug,Wands:2000dp,Lyth:2003im})  around a spatially homogeneous and isotropic background, 
the Friedmann-Lema$\hat{\i}$tre-Robertson-Walker (FLRW) space-time, 
the relevant part of the perturbed metric is given by 
\aln{
g_{\mu\nu}dx^\mu dx^\nu=   - \lambda^{2}(x,t) dt^2+   a^{2}(x,t) \delta_{ij}dx^idx^j  \label{FLRW}
}
where
$a(x,t) = \overline{a}(t) e^{\psi(x,t) }$ is the scale factor of each separate universe labeled by the ``spatial coordinate'' $x$.  Background quantities are denoted with bars such as $\overline{a}(t)$ or $\overline{\rho}(t)$.
$\lambda (x,t)$ is the lapse function which allows us to reparametrize the time coordinate of each separate universe. 
The shift vectors can be dropped as far as we focus on scalar perturbations in the long wave length limit. 

With $\overline{N}(t) :=  \ln \overline{a}(t)$, we can define
the local e-folding number as
 \aln{N(x,t) = \overline{N}(t) + \psi(x,t), \label{local e-folding number}}
 from which 
the local expansion rate is given by 
\aln{H(x,t) = \frac{1}{\lambda(x,t)} \frac{d \ln a(x,t)}{dt} = \frac{\dot{\overline{N}}(t)+ \dot{\psi}(x,t)}{\lambda(x,t)}  ~, \label{local Hubble}} where the dot denotes a $t$-derivative.
From the Einstein-Hilbert action, the Hamiltonian constraint gives $H^{2}(x,t) = \rho(x,t)/3 m_{\rm Pl}^{2}$ in the superhorizon limit, where $\rho(x,t)$ is the total energy density and $m_{\rm Pl}$ is the reduced Planck mass.

One of the important quantities in the cosmological perturbation theory is the uniform density curvature perturbation $\zeta$, and  given non-linearly in terms of the density fluctuations as
\be
-\zeta = \psi + \frac{1}{3} \int_{\bar{\rho}(t)}^{\rho(x,t)} \frac{d \rho'}{\rho' + P},
\label{non-linear zeta}
\ee
at each separate universe $x$ at time $t$. 
Here, $P(x,t)$ is the total pressure. 
%$\rho(x,t), P(x,t)$ are the total energy density and pressure with inhomogeneous fluctuations. 
Note  that one has $-\zeta = \psi$ on a uniform density slice, i.e. for $\delta\rho = \rho(x,t)-\overline{\rho}(t)=0$.  
At the linear order, this becomes the well-known  gauge-invariant expression of the curvature perturbation,
\be
-\zeta^{(1)} = \psi^{(1)} + \frac{\delta \rho}{3(\overline{\rho} + \overline{P})}  
=  \psi^{(1)}- \overline{H} \frac{ \delta \rho}{\dot{\overline{\rho}}}  ~ \label{linear-curvature-perturbation}
\ee
 where we have used the energy conservation law,
 $ \dot{\overline{\rho}}+3\overline{H}(\overline{\rho}+\overline{P})=0$, in the second equality.
Here  $\overline{H}=\dot{\overline{N}}$ and
the superscript ${}^{(1)}$ denotes the linear order.
 $\zeta$ is gauge invariant, but note that it is not conserved even on the superhorizon scale 
 unless the total energy density and pressure satisfies a barotropic equation of state (EOS) with $P=P(\rho)$. 
Thus if the energy density consists of e.g., both matter and radiation, $\zeta$ is not conserved. 

For a perfect fluid $X$ with a barotropic EOS, $P_X=\omega_X \rho_X$, which is
energetically isolated from the rest of matter and radiation, 
we can similarly define the uniform $X$-density curvature perturbation as
\aln{-\zeta_X &= \psi + \frac{1}{3} \int_{\bar{\rho}_X}^{\rho_X} \frac{dr}{r+P_X(r)} = \psi + \frac{\ln (\rho_X / \bar{\rho}_X)}{3 (1 + w_X)} ~. 
\label{zeta_X}
}
Here we have assumed that  $w_X^{}$ is constant.  
This quantity is not only gauge invariant, but also time-independent on the super-horizon 
scale.\footnote{In fact, differentiating $\zeta_X$ with respect to $t$, we have
\be
\dot{\zeta_X} = \frac{\bar{H}}{\bar{\rho}_X +\bar{P}_X }\left(\delta P_X - \frac{\dot{\bar{P}}_X }{ \dot{\bar{\rho}}_X } \delta \rho_X\right) %~~~{\rm with}~~~~ \delta P_X^{\rm na}:= \delta P_X - \frac{\dot{\bar{P}}_X }{ \dot{\bar{\rho}}_X } \delta \rho_X = 0 ~.
= \frac{\bar{H}}{\bar{\rho}_X +\bar{P}_X }\left(\delta P_X - \frac{\delta{P}_X }{ \delta {\rho}_X } \delta \rho_X\right)=0,
\ee
where we have used the barotropic EOS;
$\dot{\bar{P}}_X /\dot{\bar{\rho}}_X = \delta P_X / \delta \rho_X = \partial P_X^{}/\partial \rho_X^{} =w_X^{} ~.$
} 
The equation (\ref{zeta_X}) can be rewritten as
\aln{
&  \rho_X^{} (x,t) = \bar{\rho }_X (t)  \ e^{-3 (1 + w_X) (\zeta_X(x,t) + \psi(x,t) ) } ~ \label{rho_X}
}
which is often used in the following. 
At the linear order, $\zeta_X^{(1)}$ is given by
\be
-\zeta_X^{(1)} = \psi + \frac{\delta \rho_X}{3(\bar{\rho}_X + \bar{P}_X)}
=  \psi - \bar{H} \frac{ \delta \rho_X}{\dot{\bar{\rho}}_X}  ~. \label{linear-X-curvature-perturbation}
\ee
The curvature perturbation $\zeta^{(1)}$  is a sum of each component $\zeta_X^{(1)}$ and written as
\be 
-\zeta^{(1)} = \psi + \frac{\sum_X \delta \rho_X}{3 \sum_X (\bar{\rho}_X + \bar{P}_X)}  = - \frac{\sum_X (1 + w_X) \Omega_X \zeta_{X}^{(1)}}{(1 + w_{\rm tot})} ~ \label{linear-curvature-perturbation-perfect-fluids}
\ee
where $\omega_{\rm tot}:= \sum_X \Omega_X \omega_X$ and $\Omega_X := \bar{\rho}_X / \bar{\rho}$.

The  adiabatic perturbation is the fluctuation of the total density $\delta \rho$ with $\delta(n_X/n_{\gamma})=0$ where $\gamma$ denotes the photon.
In this case, one has $\zeta_{X} = \zeta$ for all $X$ and the curvature perturbation $\zeta$ becomes constant on large scales.
On the other hand, fluctuations $\delta(n_X/n_{\gamma})$ with $\delta \rho=0$ are called isocurvature perturbations. 
More generally, isocurvature fluctuation between specifies of $Y$ and $Z$ is defined by \cite{Malik:2002jb}
\be
S_{YZ} = 
-3 (\zeta_{Y}^{} - \zeta_{Z}^{}).
\label{linear-entropy-perturbation}
\ee
It is also gauge invariant and time independent. 
%When a species $X$ is a matter ($\omega_X=0$) and $r$ is the radiation, $S_{X\gamma}$ is equal to the $X$-isocurvature fluctuation;
Especially,  we have
\be
S_{X} := S_{X\gamma} =  \frac{\delta(n_X/n_{\gamma})}{(n_X/n_{\gamma})}=\frac{\delta n_X}{n_X} - 3 \frac{\delta T}{T}. \label{entropy-perturbation}
\ee
Then Eq. (\ref{linear-X-curvature-perturbation}) can be  expressed as a sum of the adiabatic and isocurvature fluctuations as
\be
\zeta^{(1)}_{X} =  \zeta^{(1)} +\zeta^{(1)}_{X,{\rm iso}} ~,~~~~   \zeta^{(1)}_{X,{\rm iso}}  := \frac{\sum_Y (1 + w_Y) \Omega_Y S_{YX}^{(1)}}{3(1 + w_{\rm tot})} .
\ee
Thus the isocurvature term $\zeta_{X,{\rm iso}}^{(1)}$ represents a deviation of $\zeta_X^{(1)}$ from 
the adiabatic mode $\zeta_{}^{(1)}$.

The primordial fluctuations are almost Gaussian but
non-Gaussianity can be generated during the evolution of the fluctuations. 
The non-Gaussianity parameters, $f_{\rm NL}$ or $g_{\rm NL}$, are defined by the expansion of the curvature perturbation $\zeta$ as
\aln{
\zeta= %\zeta_1^{}+\sum_{n=2}^\infty \frac{1}{n!}\zeta_n^{}:=
\zeta_{\rm G} - \frac{3}{5} f_{\rm NL}^{}\zeta_{\rm G}^2+\frac{9}{25}g_{\rm NL}^{}\zeta_{\rm G}^3+{\cal{O}}(\zeta_{\rm G}^4). 
\label{expansion of zeta}
}
The leading term $\zeta_{\rm G}$ is a Gaussian fluctuation.\footnote{
The numerical factors, $-3/5$ and $9/25$,
 come from the relation between the curvature perturbation and the gravitational potential in matter domination; $\Phi = -3\zeta/5$. 
 }
In a scenario where adiabatic perturbation is generated by isocurvature perturbation after the primordial inflation,  
significant non-Gaussianity can be produced even if the primordial fluctuation during inflation is Gaussian.
It is because 
the transfer mechanism from isocurvature to adiabatic fluctuations is non-linear as well as 
the evolution of axion field in a cosine potential. 
Consequently, $\zeta_A^{}$ has contributions from the  higher order terms of the Gaussian fluctuation $\delta A$ 
which leads to the non-Gaussianity of $\zeta$.  
See also 
%\cite{Sasaki:2006kq},\cite{ArkaniHamed:2003uz}-\cite{Komatsu:2010hc} 
\cite{Sasaki:2006kq,ArkaniHamed:2003uz,Silverstein:2003hf,Chen:2006nt,Langlois:2008vk,Byrnes:2010em,Bartolo:2010bj,Burrage:2010cu,Komatsu:2010hc} 
and references therein for non-Gaussianity in the curvaton scenarios and other types of non-Gaussianity.

%%%%%%%%%%%%%%%%%%%%%%%%%%%%%%%%%%%
%%%%%%%%%%%%%%%%%%%%%%%%%%%%%%%%%%%
%%%%%%%%%%%%%%%%%%%%%%%%%%%%%%%%%%%
%______________________________________________________
\subsection{Observational constraints} \label{sec:current status}
We now focus on fluctuations of the axion field $A$. 
The cold dark matter (CDM) is assumed to be composed of axions $A$ and other unspecified particles denoted by  $d$. 
In order to compare with the Planck 2018 observations  \cite{Akrami:2018odb},  adiabatic\footnote{
In general, the adiabatic mode is given by the curvature perturbation during an early radiation dominated era,  
${\cal R} =  -\zeta_{\rm rad}^{}$, see \cite{Gordon:2002gv}. 
In this paper, we assume that $\Omega_{\gamma} \simeq 1$ and there are no neutrino isocurvature perturbations, 
so that we have $\zeta_{\rm rad}^{} \simeq \zeta_{\gamma}$.
Also, we assume  absence of baryon and dark matter isocurvature perturbations; $\zeta_{d} = \zeta_{\gamma}$. 
%It is also called ``curvature'' perturbation since $\zeta_{r}$ dominates the curvature perturbation (\ref{LSS-linear-curvature-perturbation}) in the radiation dominated era with $\Omega_{\gamma} \simeq 1$. 
}
and isocurvature modes, ${\cal R}$ and ${\cal I}$, are introduced as
\be
{\cal R} :=  -\zeta_{\gamma}^{} ~,~~~{\cal I} :=  %\frac{3\Omega_m^{}}{\Omega_{c}^{}} \zeta_{\gamma,{\rm iso}}^{(1)}\bigg|_{\rm lss} :=\frac{3}{r_c^{}}\zeta_{\gamma,{\rm iso}}^{(1)}\bigg|_{\rm lss}\simeq  
r_{A}^{} S_{A\gamma}^{}= 3 r_A^{}(\zeta_\gamma - \zeta_{A}^{}) ~,  \label{definition of R and I}
\ee
where $r_A$ is the ratio of the abundance of the axion  to the total CDM today;
\aln{
r_A^{}:=\frac{\Omega_A^{}}{\Omega_{\rm CDM}^{}}\bigg|_{\text{today}}^{},\quad \Omega_{\rm CDM}^{}=\Omega_d^{}+\Omega_A^{}. 
}
The Fourier expansion of ${\cal{R}}$ (and also ${\cal{I}}$) is given by 
\aln{{\cal{R}}(x)=\int \frac{d^3\mathbf{k}}{(2\pi)^3}e^{-ikx}{\cal{R}}_\mathbf{k}^{},%\quad {\cal{R}}=\int \frac{d^3k}{(2\pi)^3}e^{-ikx}{\cal{I}}_\mathbf{k}^{}
}
and its  power spectrum is defined as 
\aln{
\langle {\cal{R}}_k^{}{\cal{R}}_{k'}^{}\rangle=(2\pi)^3\delta^{(3)}(k+k')P_{{\cal R}}^{}(k),\quad {\cal{P}}_{{\cal{RR}}}^{}(k):=\frac{k^3}{2\pi^2} P_{{\cal{R}}}^{}(k).
\label{definition of PRR}
}
${\cal{P}}_{{\cal RI}}^{}(k)$ and  ${\cal{P}}_{{\cal II}}^{}(k)$ are defned in a similar manner.

Planck 2018 results \cite{Akrami:2018odb} give the amplitude of the scalar perturbations, 
\aln{
A_s^{}:={\cal{P}}_{{\cal RR}}^{}(k_*^{})=2.1\times 10^{-9}, 
\label{Planck-scalar}
}
where $k_*^{}=0.05\text{Mpc}^{-1}$ is the reference (pivot) scale. 
For the slow-roll inflation in a potential $V$, the scalar perturbation is written
\aln{
{\cal{P}}_{{\cal RR}}^{}(k)=\frac{V}{24\pi^2 m_{pl}^2\epsilon}\bigg|_{k=a\overline{H}}^{} \label{adiabatic from inflation}
}
 in terms of the slow roll parameter $\epsilon =m_{\rm Pl}^2(V'/V)^2/2$. 
The magnitude of non-adiabaticity is measured by 
\be
\beta_{\rm iso}(k) = \frac{{\cal P}_{\cal II}}{{\cal P}_{\cal RR} +  {\cal P}_{\cal II}} ~~~,~~~~\cos \Delta = \frac{{\cal P}_{\cal RI}}{\sqrt{{\cal P}_{\cal RR} {\cal P}_{\cal II}}}.
\ee
The observational bounds by Planck 2018 \cite{Akrami:2018odb} are given by 
\be
\beta_{\rm iso}(k_*^{})
 < 0.038 ~~~~~{\rm for}~~~~ \cos \Delta = 0 
 \label{uncorrelated}
\ee
for ${\cal I}$ and ${\cal R}$ being uncorrelated, 
or 
\be
\beta_{\rm iso}^{}(k_*^{})
< \left\{ \begin{matrix} 0.000950 \\ 0.00107 \end{matrix} \right. ~~~~~{\rm for}~~~~ \begin{matrix} \cos \Delta = +1 \\ \cos \Delta = -1 \end{matrix}  \label{correlated}
\ee
for fully (anti-)correlated cases.
In the axion scenario we discuss below, ${\cal{P}}_{\cal RR}^{}$ and ${\cal{P}}_{\cal II}^{}$ 
has the same origin as the axion isocurvature mode in the axion primordial fluctuation $\delta A$. 
Therefore, adiabatic and axion-isocurvature fluctuations are fully anti-correlated; i.e. $\cos \Delta =-1$. 

\

Finally, observational bounds for the non-Gaussianity \cite{Akrami:2019izv} are given by 
\aln{
f_{\text{NL}}^{\rm local}=4\pm 20,%-0.9\pm 10 %5.1 
\quad g_{\rm NL}^{}= (-5.8\pm 13 %6.5
)\times 10^4,\quad (95\%\text{CL by Planck 2018}), 
\label{f and g}
}
where the superscript ``local" means that the three point function (bispectrum) 
of the Bardeen's gravitational potential $\Phi$ is given by the
following product of two point functions $P$;  
$B^{\rm local}_\Phi(k_1^{},k_2^{},k_2^{})=2f_{\rm NL}^{\rm local} \left( P_\Phi^{}(k_1^{})P_\Phi^{}(k_2^{})+\cdots \right)$
 where $P_\Phi^{}(k)$  is the power spectrum of $\Phi$.  
%\red{We can actually show this form of bispectrm by using Eq. (\ref{expansion of zeta}) 
%and the relation $\Phi=-3\zeta/5$ during matter domination. }
This category of bispectrm usually arises in multiple field inflation models or when extra light scalar fields, different from the inflaton field, contribute to the final curvature perturbation.  See e.g. \cite{Byrnes:2010em,Komatsu:2010hc} for more details.  
Axion scenario discussed in this paper also belongs to this category. 
%\mage{この$f^{\rm local}_{\rm NL}$はisocurvatureが無いとした時のものだと思うので、むしろ\cite{Akrami:2019izv}のTable 7の$f^{{\rm a,aa}}_{\rm NL}$への制限(を95\%CLに直したもの)を使うべきかと思います。ここでの記号は$f^{\rm local}_{\rm NL}$のままで大丈夫です。}

%(See Eq.(\ref{expansion non gaussianity}) for example.)
%
%See \cite{Akrami:2019izv} for more details. 

%%%%%%%%%%%%%%%%%%%%%%%%%%%%%%%%%%%
%%%%%%%%%%%%%%%%%%%%%%%%%%%%%%%%%%%
%%%%%%%%%%%%%%%%%%%%%%%%%%%%%%%%%%%
%__________________________axion scenario_____________________________
\section{Axion-CMB Scenario}\label{axion scenario}
Now we consider a possibility that the primordial axion fluctuations generate the scalar
amplitude of the CMB anisotropy. It is similar to the curvaton scenario in that
the CMB anisotropy originates in fluctuations of a field other than inflaton, 
but  a difference is that the axion field is assumed to be stable in our scenario.
Namely, we consider QCD-like axions that do not decay until present.
Thus, unlike generating the CMB anisotropy by decay of curvaton, 
fluctuations in the radiation sector is induced
when axion potential is generated, since
 the local conservation of energy between radiation and axions converts fluctuations of
 the axion to those of radiation.
In the case of the QCD-axion, this conversion occurs at the QCD phase transition. 
%But it is well known  that this axion-induced  temperature fluctuation cannot be consistent with the CMB observation
%unless a sufficient dilution of the axion energy density occurs such as thermal inflation after QCD phase transition. 

In this section, we obtain conditions for the above scenario to be consistent with the CMB observations: the amplitude of the scalar power spectrum, the isocurvature constraint
and the non-Gaussianity constraint. 
The following  parameters are constrained,
\begin{itemize}
\item primordial fluctuations of axions, $\langle (\delta A_{\rm ini}/\overline{A}_{\rm ini})^2\rangle \sim \overline{H}_{\rm exit}^2/(f_A^{}\overline{\theta}_{\rm ini})^2$
%\item ratio of energy densities $R=\Omega_A^{}/\Omega_r^{}$ right after the conversion at $T=T_A^{}$
\item $R=\Omega_A/\Omega_r$, ratio of energy densities of axion to radiation right after the potential generation
\item $r_A=\Omega_A^{}/\Omega_{\rm CDM}|_{\rm today}$, fraction   of the present axion abundance.
\end{itemize}
Here $\overline{H}_{\rm exit}$ is the background Hubble parameter of the primordial inflation when the axion field fluctuations are generated. 
$f_A$ is the axion decay constant and $\bar{\theta} \in [-\pi, \pi]$ is the misalignment angle.
%We also include \mage{quantities ${\cal X}$, ${\cal Y}$ and ${\cal Z}$ 
%which encode effects of anharmonicity of the axion potential and history of axion's evolution.}
%effects of anharmonicity of the axion potential, which are denoted by ${\cal X}_{\rm a}$ and ${\cal X}_{\rm i}$.
%It makes the misalignment angle dependence much richer than a simple harmonic approximation. 
Our goal is to investigate the allowed region of these three parameters and to construct possible particle physics models
in the next section. 
In the investigations,
we take effects of anharmonicity in the axion potential and nonlinearity in the evolutions of the axion field before and after the QCD transition, which are denoted by ${\cal X}$, ${\cal Y}$ and ${\cal Z}$, respectively.

%%%%%%%%%%%%%%%%%%%%%%%%%%%%%%%%%%%%%%%%%%%%%%%%%%%%%%
%%%%%%%%%%%%%%%%%%%%%%%%%%%%%%%%%%%%%%%%%%%%%%%%%%%%%%
%%%%%%%%%%%%%%%%%%%%%%%%%%%%%%%%%%%%%%%%%%%%%%%%%%%%%%
%__________________________Temperature fluctuation from  axion_____________________________
\subsection{Isocurvature perturbations and non-Gaussianity}
In this section, based on the formulas discussed in the previous section, 
we calculate (iso)-curvature perturbations and its non-Gaussianity.
We assume an existence of a QCD-like axion
which is massless during the primordial inflation, acquires a mass at temperature $T=T_A^{}$ and does not 
decay until present.
When axion potential is generated, 
the increase of the axion potential energy is compensated by decrease of the radiation energy.
Thus it generates isocurvature fluctuations. 
In this subsection,  the energy transfer from radiation to axion is assumed to occur instantaneously.
Gradual energy transfer is discussed in \ref{app:gradual}, but the results are not so much different.   

\begin{figure}[t!]
\begin{center}
\includegraphics[width=12cm]{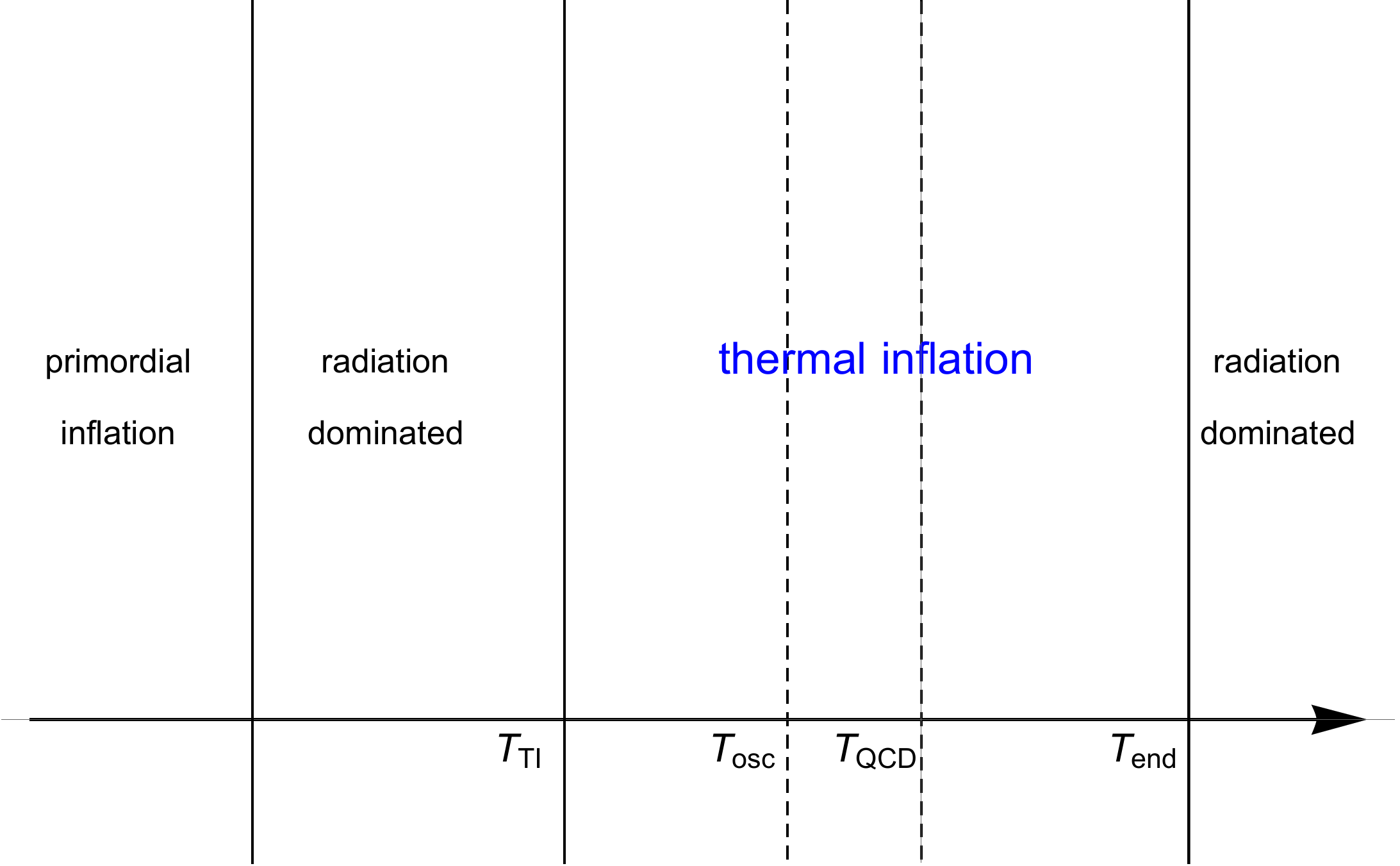}
\caption{Thermal history of (QCD) axion scenario with thermal inflation. 
The horizontal axis represents the direction of time evolution. 
Here, different boundaries are labeled by the  temperatures of background radiation. 
}
\label{fig:history}
\end{center}
\end{figure}

Thermal history of the universe can be divided into several different phases and we need to impose appropriate boundary conditions at each phase boundary.  See Fig.\ref{fig:history} for  a schematic picture of the thermal history.
In any case,  even when the energy transfer occurs instantaneously, 
the total energy density $\rho$ and the spacetime metric, especially, the scale factor $a=\exp ( \bar{N} + \psi)$ must be continuous.
On the other hand, 
 each component of the energy density $\rho_X$ can be discontinuous and consequently
$\zeta_X$ has discontinuity. 
The curvature perturbation $\zeta=-\psi +\delta \rho/3(\rho+P)$ is also discontinuous 
unless $\delta \rho =0$
since  the EOS and accordingly the pressure $P$ is discontinuous 
between adjacent phases.\footnote{
The authors in \cite{Gong:2016yyb}, based on their assumption that the total curvature perturbation $\zeta$ is continuous, concluded that CDM's density perturbation existing before and during a thermal inflation can originate the observed CMB anisotropy with suppressed isocurvature perturbation.}

\

%%%%%%%%%%%%%%%%%%%%%%%%%%%%%%%%%%%%%%%%%%%
\noindent$\bullet$ {\bf Axion perturbations and amplitude of CMB}\\
PQ symmetry of the axion field $A$ is assumed to be already broken before the primordial inflation.
Then axion field fluctuations $\delta A \neq 0$ are generated during the primordial inflation.  
In a very high temperature universe,
axion does not have potential and its energy density is negligible. 
Suppose that axion potential is  suddenly generated when temperature drops down to $T_{A}$. 
Then axion energy density  increases by transferring energy from radiation. 
By an appropriate gauge transformation, 
we can take a constant time slice on which this sudden transition takes place.
In absence of any other components to fluctuate, 
this is a uniform density slice  $\delta \rho=0$, and thus we have 
$\psi = - \zeta_{\rm inf}$ where $\zeta_{\rm inf}$ is the curvature perturbation originated from inflaton's perturbation. 
Before the transition, the density fluctuation is equal to the temperature fluctuation and the slice is characterized by $T=T_A^{}$. 
In the following, we generally call this slice ``{\it QCD slice}". 
 %\footnote{\mage{We assume that axion's perturbation $\delta A$ does not affect the transition temperature: $\delta T_{A}/\delta A = 0$. この脚注は必要？変わる可能性はなぜある？}}  
After the transition, on the other hand, radiation energy is transferred to the axion field 
which has inhomogeneous primordial fluctuations, and the slice 
is no longer a uniform temperature slice.  
Indeed, from the energy conservation right after the transition, we have
$\delta \rho_r =-\delta \rho_A \neq 0$ whose sum vanishes
\aln{
& \delta \rho 
= \delta \rho_{r}^{}  + \delta \rho^{}_{A} 
= 0 ~,
}
or equivalently,  using Eq. (\ref{rho_X}), 
\aln{
\Omega_{r}^{} \left[ e^{{-4(\zeta_{r}^{}- \zeta_{\rm inf} )}} -1 \right]+  \frac{\rho_{A} - \bar{\rho}_{A} }{\bar{\rho}}  = 0.
\label{on-QCD-slice-V}
}
Solving this equation with respect to $\zeta_{r}^{}$, we obtain
\be
 \zeta_{r}^{}  = \zeta_{\rm inf} - \frac{1}{4} \ln \left( 1 -  R^{} \frac{\rho_{A} - \bar{\rho}_{A} }{\bar{\rho}_{A}}\right) ~ 
\label{adiabatic-mode}
\ee
where
\be
R^{} := \frac{ \Omega_{A}^{}}{ \Omega_{r}^{}} \bigg|_{\rm right~after~transition} 
\label{def of R}
\ee
is the ratio of the energy densities of axion to radiation evaluated right after the sudden potential generation at $T=T_A^{}$. 

Axion energy density is given by the axion potential $\rho_{A} \simeq V_{A}(\bar{A})$
either for a slow-rolling case ($\bar{A}$ is the field value) or for an oscillating case ($\bar{A}$ is the amplitude of the oscillation). 
In our situation, we assume that $\zeta_{\rm inf}$ is too small to explain the CMB anisotropy and focus on 
the axion fluctuations as its origin. In the following we set $\zeta_{\rm inf}=0$.
At the linear order in the fluctuation of $\delta A_{\rm ini}$, $\zeta_r$ in Eq. (\ref{adiabatic-mode}) becomes
\aln{
%\zeta_{r}^{(1)} = 
\zeta_{r {\rm G}} = \frac{R}{4} \left. \frac{V_{A}^{'}}{V_{A} } \right|_{A = \bar{A}_{\rm QCD}}  \times
\left( \frac{\delta A_{\rm QCD}}{\delta A_{\rm ini}}\right) \delta A_{\rm ini} 
= {\cal X}(\bar{\theta}_{\rm QCD}) ~ {\cal Y}(\bar{\theta}_{\rm QCD})  ~ \frac{R}{2}\frac{\delta A_{\rm ini}}{\bar{A}_{\rm ini}}  ~,  
\label{scalar amplitude}
}
where we rewrote the fluctuation $\delta A_{\rm QCD}$ at $T_{\rm QCD}$ in terms of the initial Gaussian 
fluctuation $\delta A_{\rm ini}$ created during
the primordial inflation. 
$\theta = A/f_{A}$ is the dimensionless angle of the axion field. 
We also
defined anharmonicity factors ${\cal X}(\bar{\theta}_{\rm QCD} ) $ and ${\cal Y}(\bar{\theta}_{\rm QCD})$
by
\aln{
{\cal X}(\bar{\theta}_{\rm QCD} ) := \frac{1}{2}\left. \frac{\partial \ln V_{A}}{\partial \ln A} \right|_{A = \bar{A}_{\rm QCD}}   = \frac{\bar{\theta}_{\rm QCD} \sin (\bar{\theta}_{\rm QCD})}{2(1 - \cos  (\bar{\theta}_{\rm QCD}) )}= 
\frac{\bar{\theta}_{\rm QCD}}{ 2 \tan(\bar{\theta}_{\rm QCD}/2)} ~,\label{anharmonicity X}
}
\aln{ {\cal Y}(\bar{\theta}_{\rm QCD})  :=\frac{\bar{\theta}_{\rm ini}}{\bar{\theta}_{\rm QCD}} \overline{\frac{\delta \theta_{\rm QCD}}{\delta \theta_{\rm ini}}} ~, \label{anharmonicity Y}
}
${\cal X}$ is the anharmonicity factor associated with the anharmonicity of the potential $V_A(A)$, and 
we have used the explicit sinusoidal form of the potential $V_{A}\propto 1 - \cos \theta$ in Eq.~(\ref{QCD axion potential}). 
In the harmonic limit with $\theta \ll 1$ it goes to unity. 
The factor ${\cal Y}$ takes into account axion's evolution before the QCD slice. 
The overline denotes that $\theta$ is replaced by the spatially homogeneous angle $\bar{\theta}$. 
It is almost unity for the scenarios we work on in the next section; either axion's evolution is almost linear or the axion  field does not evolve so much before the QCD slice, see Eq.~(\ref{thermal inflation Y}).  
The subscript G of $\zeta_{r{\rm G}}^{}$ stands for the fact that axion's fluctuations created during the primordial inflation is Gaussian, and they have the almost scale invariant spectrum,
\aln{
\langle \delta A_{{\rm ini}}(k) \delta A_{{\rm ini}}(k') \rangle = (2\pi)^{3} \delta^{(3)}(k+k') \frac{\overline{H}^{2}_{\rm exit}(k)}{2 k^{3}} ~,~~~~ \overline{H}_\text{exit}^{}(k) := \overline{H}|_{k=a\overline{H}} ~.
\label{gaussianA}
}

\vspace{5mm}

Generally speaking, the above curvature perturbation  $\zeta_r^{}$ is not yet the final CMB fluctuation we observe today because further mixings with other fields after the transition $T=T_A^{}$ may contribute to the curvature perturbation. 
If there are no such mixings and the energy density of radiation is taken over to the current density of radiation, this gives the final CMB fluctuations. 
In thermal inflation scenario discussed in next section, as well as in the standard QCD scenario, the fluctuation Eq. (\ref{scalar amplitude}) is almost copied to the fluctuation of radiation, i.e. CMB anisotropy. 
See \ref{final fluctuation} for more details.    
Therefore, from Eq.~(\ref{scalar amplitude}), the scalar spectrum amplitude of CMB is given by 
\footnote{%\red{As for the scalar spectral index $n_s^{}\sim 0.96$, it is difficult to explain this by the fluctuation of  QCD axion as long as we consider the standard scenario of it. However, because $n_s^{}$ depends on the derivative of power spectrum, we can always realize $n_s^{}\sim 0.96$ as usual  by the fluctuation of inflaton in primordial inflation. We thank the referee for pointing this out.  }
The scale dependence of the power spectrum $A_{s}$ comes from the time dependence of the Hubble rate during the primordial inflation where axion's fluctuations are generated: $n_{s} -1  \simeq   d \ln \overline{H}^{2} /d \overline{N} |_{k_{*} = a\overline{H}} = -2 \epsilon $. This is the same scale dependence as the one for the tensor mode because of the absence of the axion potential during the inflation. It should be possible to obtain the observed spectral index $n_{s}\simeq 0.965$ by setting inflaton's potential properly. % while we keep it unspecified in this work. 
Note that the time evolution of the axion field value due to its potential, encoded in $\cal X$ and $\cal Y$, does not bring any scale dependence to $A_{s}$ since the CMB scale is far out of the horizon during axion's evolution.
}
\aln{
\sqrt{A_s} = %\sqrt{{\cal{P}}_{{\cal RR}}^{}(k_*^{})}=
R \times  \frac{\overline{H}_\text{exit}^{}(k_*^{})}{4\pi f_A \overline{\theta}_{\rm ini}} 
{\cal X}(\bar{\theta}_{\rm QCD}) {\cal Y}(\bar{\theta}_{\rm QCD})  ~.
%=\sqrt{2.1\times 10^{-9}}=4.6\times 10^{-5} 
\label{CMB by axion}
}
Comparing it with the CMB observation in Eq.~(\ref{Planck-scalar}), this gives a relation between $R$ and $\overline{H}_\text{exit}^{}(k_*^{}) / \pi f_A$ for each misalignment angle $\bar{\theta}_{\rm QCD}^{}$. 

\

%%%%%%%%%%%%%%%%%%%%%%%%%%%%%%%%%%%%%%
\noindent$\bullet$ {\bf Non-Gaussianity}\\
We then calculate the non-Gaussianity.  
With the assumption $\zeta_{\rm inf} = 0$ and $\rho_{A} \simeq V_{A}(A)$, the nonlinear expression Eq.~(\ref{adiabatic-mode}) is expanded with respect to $\delta A_{\rm QCD}$ as
\aln{
\zeta_{r}^{}
&=\frac{R}{4} \left. \frac{\bar{V}_A^{'}}{\bar{V}_A^{}} \right|_{\rm QCD}\delta A_{\rm QCD}+\frac{R}{8}\left[ \frac{\bar{V}_A^{''}}{\bar{V}_A^{}}+R\left(\frac{\bar{V}_A'}{\bar{V}_A^{}}\right)^2\right]_{\rm QCD} (\delta A_{\rm QCD} )^2  \nn \\
&~~~~~~~~~~~~~~~~~+\frac{R}{12}\left[ \frac{\bar{V}_A^{'''}}{2\bar{V}_A^{}}+\frac{3R}{2}\frac{\bar{V}_A^{'}\bar{V}_A^{''}}{\bar{V}_A^{2}}+R^2\left(\frac{\bar{V}_A'}{\bar{V}_A^{}}\right)^3\right]_{\rm QCD}  (\delta A_{\rm QCD})^3+\cdots \label{expansion non gaussianity} 
\\
& =\zeta_{r{\rm G}}+\frac{2}{R}\left[ \frac{ \bar{V}_A^{''}\bar{V}_A^{}}{ \bar{V}_A^{'2}}+R \right]_{\rm QCD} \zeta_{r{\rm G}}^{2} +\frac{16}{3R^2}\left[\frac{ \bar{V}_A^{'''} \bar{V}_A^2 }{2 \bar{V}_A^{'3}}+\frac{3R}{2}\frac{ \bar{V}_A^{''}\bar{V}_A^{} }{ \bar{V}_A^{'2}}+R^2\right]_{\rm QCD} \zeta_{r{\rm G}}^{3} +\cdots ~. \nn
}
%where \mage{$\zeta_{r{\rm G}}$} is given in Eq.(\ref{scalar amplitude}).
In the last line, $A_{\rm QCD} = A_{\rm ini}$ is assumed for brevity, so that ${\cal Y} = 1$. 
The anharmonicity factor ${\cal Y}$ gives a small correction to the non-Gaussianity, which we discuss later in Eqs.~(\ref{f(yne0)})(\ref{g(yne0)}).  
From this, we obtain
\aln{f_{\rm NL}^{}&=-\frac{10}{3R} \left[ \frac{ \bar{V}_A^{''}\bar{V}_A^{}}{ \bar{V}_A^{'2}}+R \right]_{\rm QCD}  ~,
\label{f(y=0)}\\
g_{\rm NL}^{}&=\frac{1}{3}\left(\frac{20}{3R}\right)^2 \left[\frac{ \bar{V}_A^{'''} \bar{V}_A^2 }{2 \bar{V}_A^{'3}}+\frac{3R}{2}\frac{ \bar{V}_A^{''}\bar{V}_A^{} }{ \bar{V}_A^{'2}}+R^2\right]_{\rm QCD} \label{g(y=0)} \\
&=\frac{1}{6}\left(\frac{20}{3R}\right)^2 \left. \frac{\bar{V}_A^{'''}\bar{V}_A^2}{\bar{V}_A^{'3}}\right|_{\rm QCD} +\frac{20}{3}f_{\rm NL}^{}-\frac{1}{6}\left(\frac{20}{3}\right)^2.  \nn
}
In the case of the sinusoidal potential $V_A^{}(\overline{\theta}) \propto 1-\cos\overline{\theta}$, we have
\aln{
&f_{\rm NL}^{}=-\frac{10}{3R}\frac{\cos (\overline{\theta}_{\rm QCD}) \left( 1-\cos (\overline{\theta}_{\rm QCD}) \right)}{\sin^2 (\overline{\theta}_{\rm QCD} ) } -\frac{10}{3} ~,
\\
&g_{\rm NL}^{}=-\frac{1}{6}\left(\frac{20}{3R}\right)^2%\frac{(1-\cos\theta)^2}{\sin^2\theta}
\tan^2\left(\frac{\overline{\theta}_{\rm QCD}} {2}\right)+\frac{20}{3}f_{\rm NL}^{}-\frac{1}{6}\left(\frac{20}{3}\right)^2. 
} 
Note that the non-Gaussianities are inversely proportional to $R$ since the leading Gaussian fluctuation is proportional to $R\ (<1)$ while the leading non-linear terms also proportional to $R$ instead of decreasing with higher orders of $R$. 
%We will later discuss 
%how the nonlinear evolution before the QCD slice affects them;  see (\ref{f(yne0)}) and (\ref{g(yne0)}).

%%%%%%%%%%%%%%%%%%%%%%%%%%%%%%%%%%%%%%%%%
\begin{figure}[t!]
\begin{center}
\includegraphics[width=8cm]{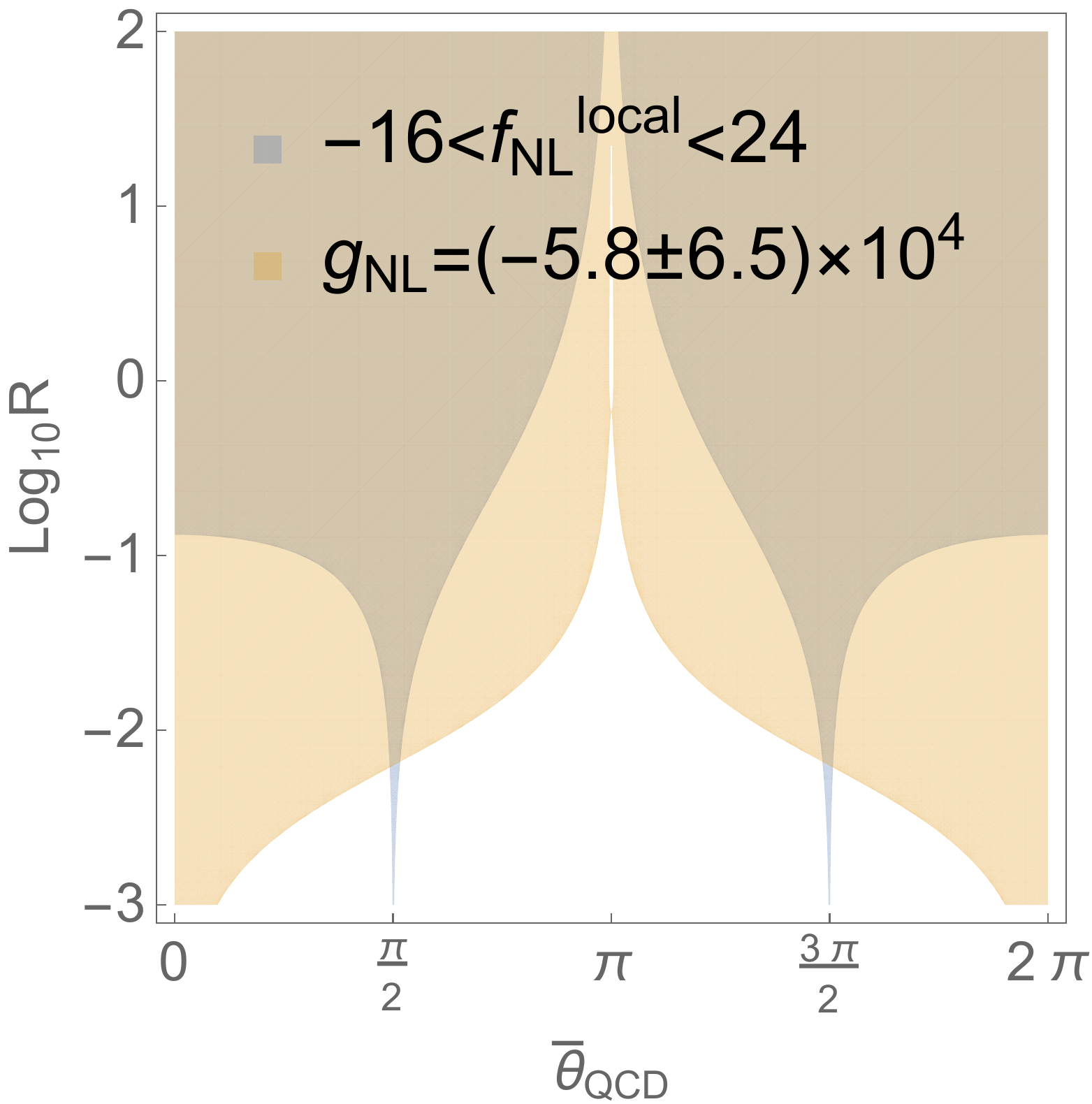}
\caption{
Allowed regions of $(\overline{\theta}_{\rm QCD},R )$ by the non-Gaussianity constraints, Eq.(\ref{f and g}). 
The blue (orange) region represents the region allowed by 
$f_{\rm NL}^{}\ (g_{\rm NL}^{})$. 
$\overline{\theta}\sim \pi/2,3\pi/2$ is necessary to obtain $R \sim {\cal{O}}(0.01)$, where $V''$ vanishes.  
}
\label{fig:NG}
\end{center}
\end{figure}
%%%%%%%%%%%%%%%%%%%%%%%%%%%%%%
In Fig.\ref{fig:NG}, we plot the allowed region of $(\overline{\theta}_{\rm QCD}, R)$ determined by the observational constraints in Eq.~(\ref{f and g}).  
The blue (orange) region corresponds to $f_{\rm NL}^{}\ (g_{\rm NL}^{})$.  
One can see that, as long as $\overline{\theta}_{\rm QCD} \sim 0$, the lower bound of $R$ is ${\cal{O}}(0.1)$, but as $\overline{\theta}_{\rm QCD} $ approaches $\pi/2$ or $3\pi/2$, the bound can be reduced to $\ {\cal{O}}(0.01)$ because $V_A^{''}$ vanishes at these points.  
It is interesting that around these values of $\overline{\theta}_{\rm QCD} $, $g_{\rm NL}$ becomes sizable while reducing $f_{\rm NL}$.

\

%%%%%%%%%%%%%%%%%%%%%%%%%%%%%%%%%%%%%
%%%%%%%%%%%%%%%%%%%%%%%%%%%%%%%%%%%%%
%%%%%%%%%%%%%%%%%%%%%%%%%%%%%%%%%%%%%
\noindent$\bullet$ {\bf Isocurvature perturbations}\\
As discussed in section \ref{sec:current status},
the magnitude of the isocurvature perturbation is measured by ${\cal{I}}=r_A^{} S_A^{}$,  where $S_A^{}$ is given by  Eq.(\ref{entropy-perturbation}).  
At sufficiently late time (after the QCD transition but before the last scattering), the axion field $A(t)$ becomes very small and we can approximate its potential by the harmonic one~\footnote{
For example, at the last scattering time, the photon temperature is ${\cal O}(0.1{\rm eV})$ and correspondingly the axion angle at this time slice is largely suppressed by a factor $(0.1{\rm eV}/T_{\rm osc})^{3/2}\sim 10^{-15}$.  
}, $V_A^{}(A) = m_{A0}^{2} A^{2}/2$, and hence, the EOS is given by $w_{A}^{} = 0$. 
By denoting such time slice as {\it late time}, and choosing an uniform temperature slice of $\zeta_r^{}=-\psi$, $S_A^{}$ can be simply evaluated to linear order in $\delta \theta$ as
\aln{
S_{A}^{} =  \ln \left. \frac{V_{A}(A)}{V_{A} (\bar{A})} \right|_{\rm late \  time} \simeq 2 \left. \frac{\delta \theta}{\bar{\theta}}\right|_{\rm late \ time} ~, \label{late time S_A}}
where we used the expression Eq.~(\ref{rho_X}) in the first equality. 
In order to relate the fluctuation at late time with the initial one of Eq.~(\ref{gaussianA}), 
we need to solve the evolution of the axion field  from the horizon exit until the late time slice.  
To quantify the power spectrum of the isocurvature fluctuation, we introduce
\aln{
{\cal Z}(\bar{\theta}_{\rm QCD}) :=  \frac{\bar{\theta}_{\rm QCD}}{\bar{\theta}|_{\rm late \ time}} \frac{\delta \theta|_{\rm late \ time}}{\delta \theta_{\rm QCD}} \label{X_i} ~.
}
The quantity ${\cal Z}$ depends  on how much energy is transferred from radiation to axion 
at the QCD slice and also on the anharmonicity of the potential in which the axion evolves after the QCD slice.
The former effect turns out to be subdominant since $R\ll 1$.
Indeed, as we will see in next section, the conventional QCD-axion scenario with the almost harmonic potential gives ${\cal Z} \simeq 1 + 3 R/4$. 
On the other hand, the thermal inflation model with QCD-axion gives 
 ${\cal Z} \simeq \bar{\theta}_{\rm QCD}/\sin (\bar{\theta}_{\rm QCD}) + {\cal O}(R)  >1$ since axion field rolls down in the anharmonic region of the potential during thermal inflation, see Eq.~(\ref{thermal inflation Z}).

Now the isocurvature power spectrum is given as 
\aln{
{\cal{P}}_{{\cal II}}^{}(k)=\frac{k^3}{2\pi^2}P_{{\cal{I}}}^{}(k)=
\left( {\cal Y}(\bar{\theta}_{\rm QCD}) {\cal Z}(\bar{\theta}_{\rm QCD})~\frac{r_A^{}\overline{H}_{\rm exit}(k)}{\pi f_A \overline{\theta}_{\rm ini}}\right)^2 ~.
}
Note that  %${\cal Z}$ multiplied by ${\cal Y}$ in (\ref{scalar amplitude}) 
${\cal Y}{\cal Z}=(\bar{\theta}_{\rm ini} / \bar{\theta}|_{\rm late \ time} )  ( \delta \theta|_{\rm late \ time} / \delta \theta_{\rm ini} )$
gives an anharmonicity factor associated with the evolution of the axion field from the initial slice to the late time. 
By plugging this into Eq.~(\ref{correlated}) and using ${\cal{P}}_{{\cal RR}}^{}(k_*^{})=2.1\times 10^{-9}$, we obtain the following constraint 
\aln{
{\cal Y}(\bar{\theta}_{\rm QCD}) {\cal Z}(\bar{\theta}_{\rm QCD})~ \frac{r_A^{}\overline{H}_{\rm exit}^{}(k_*^{})}{\pi f_A^{} \overline{\theta}_{\rm ini}}<1.5\times 10^{-6}
\quad \text{ for } \cos\Delta=-1\ . 
\label{isocurvature constraint}
}
This bound gives a  strong constraint on  $r_A^{}$ and $\overline{H}_\text{exit}^{}(k)/f_A^{}$.   

\

%%%%%%%%%%%%%%%%%%%%%%%%%%%%%%%%%%%%%%%
%%%%%%%%%%%%%%%%%%%%%%%%%%%%%%%%%%%%%%%
%_____________________________________________________________
\subsection{Allowed parameter region}\label{parameter region}
\begin{figure}[t!]
\begin{center}
\includegraphics[width=12cm]{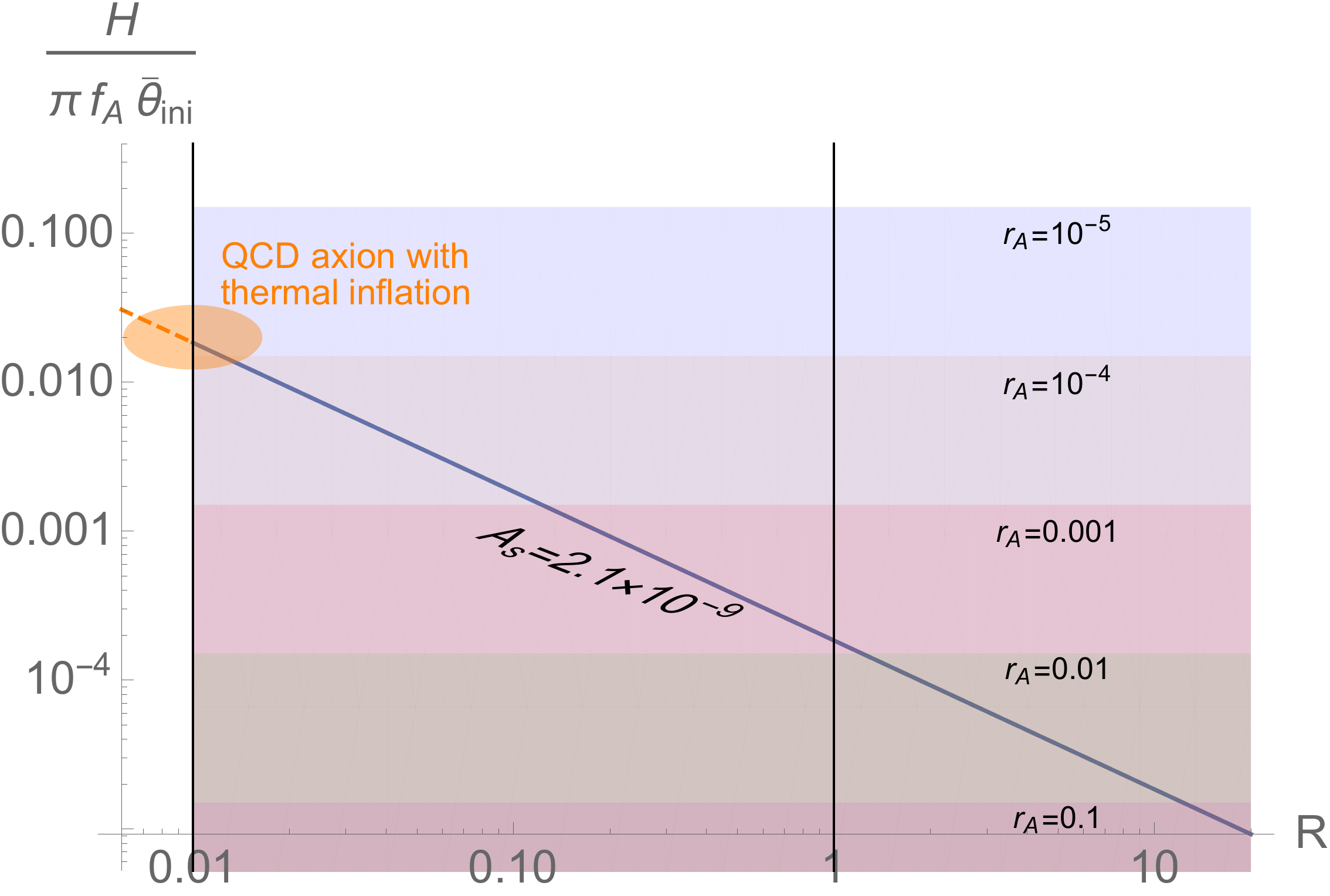}
\caption{
The parameters must be on the blue line on which the correct  scalar amplitude Eq.~(\ref{condition 1}) is produced.  
In each colored region, the maximum value of $r_A$ is indicated which is determined by Eq.~(\ref{condition 2}). 
Thermal inflation scenario discussed in the next section typically lies on the orange region. 
}
\label{fig:region}
\end{center}
\end{figure}

In the previous section, without 
specifying  models,  we calculated the (iso-)curvature perturbations and their non-Gaussianity
under an assumption that primordial fluctuations of 
QCD-axion like particles generate the CMB anisotropy.    
The results are summarized as follows:
\begin{itemize}
\item Amplitude of scalar power spectrum Eq. (\ref{CMB by axion}) is given by
\aln{\sqrt{A_s^{}}=
R {\cal X}(\overline{\theta}_{\rm QCD}) {\cal Y}(\overline{\theta}_{\rm QCD})  \frac{\overline{H}_\text{exit}^{}(k_*^{})}{4\pi f_A^{} \overline{\theta}_{\rm ini} }%=\sqrt{2.1\times 10^{-9}}
=4.6\times 10^{-5} 
\label{condition 1}
}
\item Isocurvature constraint of Eq. (\ref{isocurvature constraint}) must be satisfied,
\aln{
{\cal Y}(\overline{\theta}_{\rm QCD}) {\cal Z}(\overline{\theta}_{\rm QCD})  ~\frac{r_A^{}\overline{H}_{\rm exit}^{}(k_*^{})}{\pi f_A^{} \overline{\theta}_{\rm ini}} <1.5\times 10^{-6}
\label{condition 2}
}
\item Non-Gaussianity constraints of $f_{\rm NL}^{}$ and $g_{\rm NL}^{}$ in Fig.\ref{fig:NG}
must be satisfied. 
For most values of $\bar{\theta}_{\rm QCD}$, the ratio $R$ must satisfy the condition, $R\gtrsim 0.1$.  %Eq. (\ref{constraint on R}):
If $\bar{\theta}_{\rm QCD}$ is tuned around special values, the constraint is weakened to
\aln{R%|_{T=T_{A}^{}}
\gtrsim 0.01\quad \left(\text{around}\  \bar{\theta}_{\rm QCD}=\pi/2 \ \text{or} \ 3\pi/2  \right). 
\label{condition 3}
}
\end{itemize}
These relations (\ref{condition 1}), (\ref{condition 2}) and (\ref{condition 3}) are plotted in Fig.\ref{fig:region}.  
Here, we put ${\cal X}={\cal Y}={\cal Z}=1$ for simplicity. 

\vspace{5mm}

From Eq.~(\ref{condition 3}) and Eq.~(\ref{condition 1}), we have a condition for the Hubble parameter of the primordial inflation as
\aln{
{\cal X}(\overline{\theta}_{\rm QCD}) {\cal Y}(\overline{\theta}_{\rm QCD})  \frac{\overline{H}_\text{exit}^{}(k_*^{})}{4\pi f_A^{} \overline{\theta}_{\rm ini}} < 4.6 \times 10^{-3}.
}
Further, by eliminating
$\overline{H}_\text{exit}^{}(k_*^{})/(f_A^{}\overline{\theta}_{\rm ini})$ from Eqs.~(\ref{condition 1}) and (\ref{condition 2}),  we obtain an inequality between $r_A^{}$ and $R$ as 
\aln{
r_A^{}<
8.2\times 10^{-3}\ R \frac{{\cal X} (\overline{\theta}_{\rm QCD})}{{\cal Z}(\overline{\theta}_{\rm QCD})}~. 
\label{r and R}
}
This shows that $r_A^{} \ll 1$ and axions cannot dominate the dark matter in the current scenario. 
If the universe is radiation dominated right after the transition, we have $R\simeq \Omega_A^{}|_{T=T_A^{}}$. 
Neglecting the anhamonicity and nonlinearity factors, $r_A^{}\lesssim 8.2 \times 10^{-3}
\Omega_A^{}|_{T=T_A^{}}\ll 8.2\times 10^{-3}$ is obtained. 
As we see in next section, this rules out the standard QCD axion to explain the CMB anisotropy.

Finally, we should emphasize that the current axion scenario can predict a large value of $|g_{\rm NL}^{}|\sim 10^4$ when $R\sim 0.01$ and  $\bar{\theta}_{\rm QCD}\simeq \pi/2,3\pi/2$. 
Such a parameter region is  interesting since the scenario can be testable in the future cosmological observations of $g_{\rm NL}^{}$. 
%
%In other words, an axion model with small $R$ predicts large $g_{\rm NL}^{}$. 
%
As we will see below, a thermal inflation model with the QCD axion can be in this parameter  region. %
%(See also Fig.\ref{fig:region}.)

%%%%%%%%%%%%%%%%%%%%%%%%%%%%%%%%%%%%%%%
%%%%%%%%%%%%%%%%%%%%%%%%%%%%%%%%%%%%%%%
%%%%%%%%%%%%%%%%%%%%%%%%%%%%%%%%%%%%%%%
%%%%%%%%%%%%%%%%%%%%%%%%%%%%%%%%%%%%%%%
%________________________________________________________
\section{Models}
\label{sec:model}
Now we investigate a few particle physics models of the axion scenario for the CMB anisotropy. 
In section \ref{standard QCD axion}, we first briefly summarize why the standard QCD axion in the standard thermal history of the universe cannot satisfy the conditions; main difficulty is to satisfy the inequality Eq.~(\ref{r and R}) since axions cannot be sufficiently diluted after the QCD transition.  
Thus we need an additional mechanism to dilute the axion abundance. 
For this purpose, we consider a thermal inflation scenario in the next section \ref{thermal inflation}.  
Suppose that the temperature of the universe decreases below the QCD temperature during the thermal inflation as depicted in Figure \ref{fig:history}, one may suspect that axions are diluted even before $T_{\rm QCD}^{}$ and  $R$ becomes too small.  
We will see, however, that since the Hubble of the thermal inflation is larger than the axion mass, $3H_{\rm TI} \gtrsim m_A(T)$, the axion field does not evolve so much before $T_{\rm QCD}^{}$. 
As a result, we can realize $R\sim 0.01$ simultaneously with a small $r_A^{}$ provided that the thermal inflation lasts long enough below $T_{\rm QCD}^{}$.   
We investigate parameter regions in which the above conditions are satisfied and the axion field fluctuations can explain the CMB anisotropy. 
As a particle physics model to realize a thermal inflation, we briefly comment on a classically conformal $B$-$L$ model with a QCD axion 
%\cite{Iso:2009ss}-\cite{Iso:2017uuu}
\cite{Iso:2009ss,Iso:2009nw,Iso:2012jn,Iso:2017uuu}
 in the final section. 
A novel feature of the model is that the universe has experienced supercooling era of the $B$-$L$ and EW symmetries, and thermal inflation  occurs at around TeV scale and continues down to QCD temperature. 
However, as shown in \cite{Iso:2017uuu}, this model predicts a small Higgs vacuum expectation value  of the QCD scale, $\langle h \rangle|_{T_{\rm QCD}} \sim \Lambda_{\rm QCD}$ when the axion potential is generated, and the height of the axion potential becomes too small for getting a sufficiently large value of $R$.  
Thus we need some modifications of the original classically conformal $B$-$L$ model, such as including a Higgs-axion mixing.

%%%%%%%%%%%%%%%%%%%%%%%%%%%%%%%%%%%%%%
%%%%%%%%%%%%%%%%%%%%%%%%%%%%%%%%%%%%%%
%________________________________________________________
\subsection{No-go for the standard QCD axion}\label{standard QCD axion}
We recap the calculations in the standard QCD axion scenario to recall the difficulty of 
realizing the large-scale fluctuations.     
Below the QCD temperature $T_{\rm QCD}^{}  \simeq 150 {\rm MeV}$, the axion potential is given by
\be
V_{A0}^{} =m_{A0}^2 f_{A}^{2} [1 - \cos (A/f_{A})] ~,
\ee
\be
m_{A0} = \frac{\sqrt{m_{u}/m_{d}}}{1 + m_{u}/m_{d}} \frac{m_{\pi} f_{\pi}}{f_{A}}  \simeq  6 \times 10^{-6} {\rm eV} \times \frac{10^{12} {\rm GeV}}{f_{A}}  
\label{m_A0}
\ee
with $m_{u}/m_{d} \simeq 0.48$, $m_{\pi} \simeq 135{\rm MeV}$ and $f_{\pi} \simeq 93{\rm MeV}$. 
For $T \geq T_{\rm QCD}$, it has a strong temperature dependence
\be
V_{A}^{} 
= m_{A}(T) ^2 f_{A}^{2} [1 - \cos (A/f_{A})]
\ \text{ where } \ m_{A}^{}(T) = m_{A0}  [T_{\rm QCD}/T]^{4b} ~.
\label{QCD axion potential}
\ee
The exponent is given by $b\simeq 1.02$ in the case of three light quarks \cite{Borsanyi:2016ksw}. 
The axion acquires tiny but finite potential energy once the EW symmetry is broken at $T=T_{\rm EW} \sim 160{\rm GeV}$.  
In this work, we do not take into account the temperature dependence of the number of dynamical quarks and simply assume that $b$ is constant for $T_{\rm EW} \geq T \geq T_{\rm QCD}$. 

%%%%%%%%%%%%%%%%%%%%%%%%%%%%%%%%
\vspace{5mm}
\noindent$\bullet$ {\bf Estimation of $R$ and scalar amplitude}\\
We assume for simplicity that, the axion field value is sufficiently small and
 anharmonicity of the potential can be neglected, $V_{A} \simeq m_{A}^{2} A^{2}/2$. 
When the condition $m_{A} \geq 3 \overline{H}$ is satisfies, axion field starts oscillating.\footnote{For simplicity, 
we neglect an evolution before the oscillation. Especially when the initial angle is in the vicinity of the hilltop 
at $\theta = \pm \pi$, the nonlinear evolution of the angle becomes important in evaluating not only the axion dark matter abundance 
but also the isocurvature non-Gaussianity, see \cite{Kobayashi:2013nva} for a semi-analytical computation.}
If the oscillation occurs before $T_{\rm QCD}$, the oscillation temperature is given by
\be
T_{\rm osc}^{} &=& T_{\rm QCD} \times \left[ \sqrt{\frac{10}{\pi^{2} g_{\rm osc}}} \frac{m_{A0} m_{Pl}^{}} {T_{\rm QCD}^{2}}  \right]^{\frac{1}{4b + 2}} \label{T_osc} \\ 
&\simeq & {\rm GeV} \times \left( \frac{T_{\rm QCD}}{150 {\rm MeV}}\right)^{0.67} \left( \frac{10^{12} {\rm GeV}}{f_{A}}   \right)^{0.16}.\ 
 \notag 
\ee 
Here we used the effective number of degrees of freedom $g_{\rm osc}^{}=62$ at $T\sim {\rm GeV}^{}$ \cite{Tanabashi:2018oca} and $b\simeq 1.02$.  
Below $T_{\rm osc}^{}$,  the evolution of the axion field is almost adiabatic
and axion's ``number density'' is given by
\be
\bar{n}_{A}^{} = \bar{n}_{A}|_{\rm osc}^{} \times (\bar{a}_{\rm osc}/\bar{a})^{3} ~~~,~~~~ n_{A}|_{\rm osc} = \left. \frac{\bar{\rho}_{A}^{}}{m_{A}^{}} \right|_{\rm osc}^{} ~. \label{n_A}
\ee
Then the temperature dependence of the energy density is evaluated as 
\be
\bar{\rho}_{A}^{} &=&  m_{A} \times \bar{n}_{A}^{}  = \left( \frac{T_{\rm QCD}}{T}\right)^{4 b -3}   \left( \frac{T_{\rm QCD}}{T_{\rm osc}}\right)^{4 b + 3} V_{A0}( \bar{A}_{\rm osc})  ~. \label{rho_A}^{}
\ee

On the other hand, the ratio of the energy densities at $T=T_{\rm QCD}^{}$ is given by
\be
R 
 &=& \frac{\bar{\rho}_{A}^{}}{\bar{\rho}_{r}^{}} \bigg|_{T=T_{\rm QCD}^{}}^{} \notag 
=  \frac{m_{u}/m_{d}}{(1+ m_{u}/m_{d})^{2}} \frac{m_{\pi}^{2} f_{\pi}^{2}}{T_{\rm QCD}^{4}} \frac{30}{\pi^{2} g_{\rm QCD}}  \left( \frac{T_{\rm QCD}}{T_{\rm osc}}\right)^{4 b +3}  \frac{\bar{\theta}^{2}_{\rm osc}}{2} 
\notag \\
&\simeq & 1.2 \times 10^{-8}  \left( \frac{f_{A}^{}}{10^{12} {\rm GeV}} \right)^{1.16} \left( \frac{150 {\rm MeV}}{T_{\rm QCD}^{}} \right)^{1.67}  \bar{\theta}^{2}_{\rm osc} \label{R_QCD}^{},
\label{R in standard QCD axion}
\ee
where we used $g_{\rm QCD}^{} =69/4$ as the effective number of degrees of freedom right after the QCD phase transition to which pions also contribute.  
The axion angle $\overline\theta_{\rm osc}^{}$ can be identified with $\overline\theta=\overline\theta_{\rm ini}^{}$ because the amplitude does not change during this period.
The small numerical coefficient mainly comes from the damping of the 
oscillation amplitude, $(\bar{\theta}_{\rm QCD}/\bar{\theta}_{\rm osc})^{2} \simeq (T_{\rm QCD}/T_{\rm osc})^{4b+3} \sim 10^{-6}$. 
The smallness of $R$  already rejects the scenario, but let us go on for comparison with thermal inflation models discussed in the next section. 
We plug the result  into Eq. (\ref{CMB by axion}) together with the approximations, ${\cal X} = 1$ (harmonic approximation) and 
${\cal Y}=1$.\footnote{\label{standard Y} 
Since we neglect the gradual energy transfer from radiation to axion, the factor ${\cal Y} $ is computed as follows. 
Employing the sudden transition approximation, axion's EOS is given by $w= P/\rho= -1$ before the oscillation slice. 
Accordingly, axion's field value does not evolve, and thus, $\theta_{\rm ini} = \theta_{\rm osc}$.
As discussed in \ref{app:gradual}, 
the relation $\delta (m_{A}/3 H) =0$ on  the oscillation slice leads to the relation Eq.(\ref{on-osc-slice-III}) between  the temperature perturbations $\delta \rho_r^{}$ and $\delta \rho_{A}|_{\rm osc}^{}$. 
It causes perturbations of the e-folding number between the oscillation slice and the uniform temperature slice at $T=T_{\rm QCD}^{}$, 
which is given by $\delta N |^{\rm QCD}_{\rm osc} \simeq -(\delta \theta_{\rm osc} /\bar{\theta}_{\rm osc}) \times R_{\rm osc}/(4b +2)$ where $R_{\rm osc} := [ \rho_{A} / \rho_{r} ]_{\rm osc} \ll 1$. 
Since $n_{A}\propto \theta^{2} \propto a^{-3}$ after the oscillation slice, we obtain $\delta \theta_{\rm QCD} /\bar{\theta}_{\rm QCD} \simeq \delta \theta_{\rm osc} /\bar{\theta}_{\rm osc} - 3 \delta N |^{\rm QCD}_{\rm osc}/2$.  Therefore, we have 
${\cal Y} \simeq 1+(3 R_{\rm osc} /4)/(2b + 1) \simeq 1$.
}
Then, using the observational value of the CMB anisotropy, $A_{s}^{}\simeq 2.1 \times 10^{-9}$, we have the condition 
\aln{
 \overline{\theta}_{\rm ini}^{} \times \frac{\overline{H}_\text{exit}^{}(k_*^{})}{\pi f_A^{}}\simeq 1.6\times 10^{4} \left( \frac{10^{12} {\rm GeV}}{f_{A}^{}} \right)^{1.16} \left( \frac{T_{\rm QCD}^{}}{150 {\rm MeV}}\right)^{1.67} . \label{Condition 1 in standard QCD axion}
}

\vspace{5mm}
\noindent$\bullet$ {\bf Isocurvature fluctuations}\\
Now, let us evaluate the amplitude of the isocurvature fluctuation. 
%Because of the harmonicity, there is no nonlinear evolution after the transition. However, 
The time slice of axion's potential generation is no longer a uniform temperature slice after the  QCD transition.
Consequently, the ratio $\delta A/\bar{A}$ evaluated on a uniform temperature slice after the transition is different from the one evaluated on a uniform temperature slice before the transition.
This effect is taken into account by a factor ${\cal Z}=  1 + 3 R/4 $ which is practically irrelevant due to the
smallness of $R$.\footnote{\label{standard Z}
To see this, as in footnote \ref{standard Y}, one can compute the perturbations of the e-folding number 
between the QCD slice and a uniform temperature slice at late time,  induced by axion's perturbations on the QCD slice. 
Since $[ \delta \rho_{r} + \delta \rho_{A} ]_{\rm QCD} =0$ and the temperature dependence of axion's potential disappears on the QCD slice, we simply have $\delta N |^{\rm late}_{\rm QCD} \simeq -(\delta \theta_{\rm QCD} /\bar{\theta}_{\rm QCD} ) \times (R/2) $. 
Then, we obtain $\delta \theta_{\rm late} /\bar{\theta}_{\rm late} \simeq \delta \theta_{\rm QCD} /\bar{\theta}_{\rm QCD} - 3 \delta N |^{\rm late}_{\rm QCD}/2$, and hence, ${\cal Z} \simeq 1+3 R /4$. Although the ratio $R$ is much larger than $R_{\rm osc}$ in $\cal Y$, it is still negligibly small in ${\cal Z}$.
}
Noting that $R$ can be also rewritten in terms of the present axion abundance $r_A^{}$ as 
\aln{
R &=\frac{\rho_A^{}}{\rho_\gamma^{}}\bigg|_{\rm today}^{}\frac{(a_{\rm today}^{}/a_{\rm QCD})^3}{(T_{\rm QCD}^{}/T_0^{})^3g_{\rm QCD}^{}/g_\gamma^{}}
\nonumber
\\
%\left(\frac{\Omega_{A}^{}}{\Omega_{\rm CDM}} \frac{\Omega_{\rm CDM}}{\Omega_{\gamma}} \right)\bigg|_{\rm today}\frac{(a_{\rm today}^{}/a_{\rm QCD})^3}{}  
&= r_{A} \left.
\frac{\Omega_{\rm CDM}}{\Omega_{\gamma}} \right|_{\rm today} 
\frac{T_{0}}{T_{\rm QCD}}
\simeq 8 \times 10^{-9} \times  \ r_{A}  \left( \frac{150 {\rm MeV}}{T_{\rm QCD}} \right) 
\label{R in standard scenario}
}
where $T_0^{} \simeq 2.35 \times 10^{-4}$eV is the CMB temperature, the CDM and photon energy density fractions today are $\Omega_{\rm CDM} \simeq 2.65 \times 10^{-1}$ and $\Omega_{\gamma} \simeq 5.38 \times 10^{-5}$, respectively.
Using Eq.~(\ref{R in standard QCD axion}),
the ratio $r_A^{}$ and the misalignment angle $\bar\theta_{\rm ini}$ are related as
\aln{ 
r_{A}^{} \simeq 1.5 \times  \left( \frac{150 {\rm MeV}} {T_{\rm QCD}}\right)^{0.67} \left( \frac{f_{A}}{10^{12} {\rm GeV}} \right)^{1.16} \bar{\theta}_{\rm ini }^{2} ~. \label{theta_exit-r_A} 
}
%\blue{
%\aln{
% \bar{\theta}_{\rm ini }
%\simeq 0.3 \times \left( \frac{T_{\rm QCD}}{150 {\rm MeV}} \right)^{0.34}   \left( \frac{10^{12} {\rm GeV}}{f_{A}} \right)^{0.58} {\sqrt{r_{A}} 
% ~. \label{theta_exit-r_A}
%}
%}
%}
Then the condition (\ref{condition 2}) for the isocurvature constraint becomes 
\aln{
\overline{\theta}_{\rm ini}^{} \ \frac{\overline{H}_\text{exit}^{}(k_*^{})}{\pi f_A^{}} < 10^{-6} \left( \frac{10^{12} {\rm GeV}}{f_{A}^{}} \right)^{1.16} \left( \frac{T_{\rm QCD}^{}}{150 {\rm MeV}}\right)^{0.67}, 
}
which conflicts with Eq.~(\ref{Condition 1 in standard QCD axion}). 
In another word, the inequality Eq.~(\ref{r and R}) does not hold.

\vspace{10mm}
In this standard  scenario, there are essentially two difficulties to realize the scenario of QCD axion as the source of the CMB anisotropy. 
One difficulty is absence of sufficient dilution after the axion potential is produced until present, necessary 
to suppress the large isocurvature  fluctuations. 
In addition, since the axion potential is generated at higher temperature than $T_{\rm QCD}^{}$, the amplitude of the axion oscillation damps before the efficient energy transfer between radiation and axion at the QCD transition.\footnote{The effect of the gradual energy transfer before the transition turns out to be negligible, see \ref{app:gradual}}
Combined with the smallness of the axion potential at $T=T_{\rm osc}^{}$, these effects result in too small $R$, and it is difficult to reproduce the sufficient amount of the scalar amplitude. 
It also contradicts with the non-Gaussianity constraint. 
In the next section, we consider thermal inflation scenario in which both of the difficulties can be evaded.

%%%%%%%%%%%%%%%%%%%%%%%%%%%%%%%%%%%%%%%%%%%%%%%
%%%%%%%%%%%%%%%%%%%%%%%%%%%%%%%%%%%%%%%%%%%%%%%
%%%%%%%%%%%%%%%%%%%%%%%%%%%%%%%%%%%%%%%%%%%%%%%
%_____________________________________ 
\subsection{QCD axion with thermal inflation}\label{thermal inflation}
One of  possibilities to realize the dilution of the QCD axion field is thermal inflation
% \cite{Lyth:1995ka}-\cite{Baratella:2018pxi} 
 \cite{Lyth:1995ka,Gong:2016yyb,Hambye:2018qjv,Baratella:2018pxi} 
 that lasts until below the QCD phase transition.
The scalar field whose potential energy drives this short inflation is often dubbed ``flaton'' and 
trapped at a symmetry-enhancing point due to the thermal effect. 
Precise cosmological predictions are model-dependent. Our purpose here is 
to give a general quantitative argument of the QCD axion scenario which undergoes the thermal inflation.  
In addition to the existence of a QCD axion, we assume the following situations:    
\begin{itemize}
\item Thermal inflation starts at temperature $T_{\rm TI}^{} \gg T_{\rm QCD}^{}$.  
\item Thermal inflation ends at a low temperate $T_{\rm end}^{}\ll T_{\rm QCD}^{}$, and 
then reheating  occurs by  decay of flaton into the SM particles. 
\item The reheating temperature is higher than the oscillation temperature $T_{\rm osc}$. Subsequently, the standard Big Bang thermal history with the axion field follows as discussed in the previous section.
\item Flaton sector does not have  interactions with the QCD axion. 
\end{itemize}
%_____________________________________ 
%
The last two are assumed for simplicity. 
%
%As discussed in  \ref{final fluctuation}, the fluctuation $\zeta_r$ generated at $T=T_{\rm QCD}$ during the thermal inflation is copied as it is to the CMB anisotropy that we observe today. 
The temperature fluctuation $\zeta_r$ generated at $T=T_{\rm QCD}$ during the thermal inflation causes the fluctuation of e-foldings between the QCD slice and the ``end'' slice at $T=T_{\rm end}$ (see Eq. (\ref{''delta N''}) below), and hence, the CMB anisotropy that we observe today. One can also track the fluctuation being transferred to flaton's oscillation and then to the final radiation component, as discussed in \ref{final fluctuation}. 

The thermal inflation starts when the vacuum energy  dominates. 
Provided that the radiation is dominant before the thermal inflation, the temperature $T_{\rm TI}$ at its onset is evaluated by $V_{\rm TI}^{} = (g_{\rm TI} \pi^{2} / 30) T_{\rm TI}^{4}$ where $g_{\rm TI}^{}$ is the degrees of freedom at $T=T_{\rm TI}^{}$. 
Once the thermal inflation starts, the radiation is quickly diluted so that the Hubble expansion rate is well approximated by the constant
\aln{
H_{\rm TI}^2 =\frac{V_{\rm TI}^{}}{3 m_{Pl}^2}=(1.4\times 10^{-3}{\rm eV})^2\left(\frac{g_{\rm TI}^{}}{100}\right)\left(\frac{T_{\rm TI}^{}}{1{\rm TeV}}\right)^4 ~. \label{H_TI} } 
In the following, we treat $T_{\rm TI}^{}$, $T_{\rm end}^{}$  and the decay constant of axion $f_A^{}$
as free  parameters.

%%%%%%%%%%%%%%%%%%%%%%%%%
\vspace{5mm}

\noindent$\bullet$ {\bf Estimation of $R$ and amplitude of CMB}\\
If the axion potential is generated during the thermal inflation with the Hubble $H_{\rm TI}^{}$, the axion field does not oscillate to get diluted before the QCD transition, 
unlike the standard QCD axion discussed in the previous section.  
In order to estimate the value of $R$, 
we first solve the equation of motion (EOM) of the QCD axion field
$(-g)^{-1/2} \partial_{\mu} \left\{ (-g)^{1/2} g^{\mu\nu} \partial_{\nu} A \right\} = \partial_{A} V_{A}$ for $T \geq T_{\rm QCD}^{}$ where the potential has the strong temperature dependence Eq.~(\ref{QCD axion potential}). 
In order to avoid  dilution of axion's energy density,
we focus on the parameter region where the axion field obeys a slow-roll attractor equation 
\aln{ 3 \alpha H  \dot{A} /\lambda + \partial_{A} V_{A} =0
\label{attractor1} }
 with a numerical parameter $\alpha$.
Plugging this into axion's EOM \cite{Chiba:2009sj,Kawasaki:2011pd,Kobayashi:2013nva} and approximating the Hubble expansion rate by the constant $H_{\rm TI}^{}$, we get the condition for the attractor equation to be valid;
\aln{\frac{\partial_{A}^{2} V_{A}}{3 \alpha H_{\rm TI}^{2}} = \frac{\eta}{\alpha} \times \left( \frac{T_{\rm QCD}}{T}\right)^{8b} \cos (\theta ) 
\ll 1
}
where $\alpha = 1 + 8 b/3 \simeq 3.7 $ and
$\eta$ is given by
\aln{\eta := \frac{1}{3}\left(\frac{m_{A0}^{}}{H_{\rm TI}^{}}\right)^2 
=6.6\times 10^{-6}\left(\frac{100}{g_{\rm TI}^{}}\right)\left(\frac{10^{12}{\rm GeV}}{f_A^{}}\right)^2
\left(\frac{1{\rm TeV}}{T_{\rm TI}^{}}\right)^{4}  ~.}
The above inequality is satisfied all the way down to $T=T_{\rm QCD}$ regardless of the field value $\theta = A/f_{A}$ as far as
the  inequality $\eta \ll \alpha $ is satisfied. 
Let us define  the following evolution parameter $\Delta(t)$ by 
\aln{ \Delta(t) := H_{\rm TI} \int_{t_{\rm QCD}}^{t} \lambda dt \leq 0
\label{Delta-before}
}
 in the constant $H_{\rm TI}^{}$. 
Note that it takes a negative value since we are interested in the evolution for
$t<t_{\rm QCD}^{}$; thus it is a {\it minus} e-folding number. 
Neglecting an effect of gradual energy transfer from radiation to axion as in the previous section, we have $T/T_{\rm QCD} = e^{-\Delta }$ for $T< T_{\rm EW}$.
Then, the attractor equation, Eq.(\ref{attractor1}), is written as
\aln{\frac{d\theta}{d\Delta} = - \frac{\eta}{\alpha} e^{8 b \Delta } \sin\theta ~.  \label{SR eom of axion 1}}
Assuming $T_{\rm TI} >T_{\rm EW}$ and integrating $\Delta$ from $\Delta_{\rm EW} := \ln  T_{\rm QCD}/T_{\rm EW} \ll -1$ to $0$, we obtain
\aln{ \ln \frac{\tan \left(\theta_{\rm QCD}/2 \right)}{\tan \left( \theta_{\rm ini}/2 \right)} =   -\frac{\eta}{8b \alpha} \left( 1-e^{8b\Delta_{\rm EW} } \right)  \simeq  -\frac{\eta}{8b \alpha}  ~, \label{ini-QCD}} 
where we have plugged $\theta |_{T=T_{\rm EW}} = \theta_{\rm ini}$ and used $e^{8b \Delta_{\rm EW}}\ll 1$.\footnote{On the EW slice, there is a sudden energy transfer from radiation to axion field. Its effect on $\theta$'s perturbation is similar to the factor $\cal Z$ in Section \ref{standard QCD axion}, see footnote \ref{standard Z}, but with much smaller ratio $\Omega_{A}/\Omega_{r}$ at $T=T_{\rm EW}$ and much weaker time dependence of axion's field value.} 
We can see that, because of the strong temperature dependence of the potential $\propto T^{-8b}$, the integration is dominated by the contribution around the upper limit $\Delta =0$. 
This is also the case for $T_{\rm TI} < T_{\rm EW}$ as far as $T_{\rm TI} \gg T_{\rm QCD}$. 
Hence, we obtain 
\aln{ \theta_{\rm ini}  = 2 \arctan \left\{ e^{\frac{\eta}{8b \alpha}} \tan \left( \theta_{\rm QCD}/2 \right) \right\} = \theta_{\rm QCD} + y \sin (\theta_{\rm QCD} ) + {\cal O}\left( y^{2} \right) ~.    \label{thetaQCD}  }
On the last equality, the dependence on $y := \eta /8b \alpha \ll 1$ is expanded. 
By differentiating it with respect to $\theta_{\rm QCD}$,  
we  get
\aln{ \frac{\delta \theta_{\rm ini} }{\delta \theta_{\rm QCD} } 
 = \frac{1}{\cosh (y) - \sinh (y ) \cos (\theta_{\rm QCD})} = 1 + y \cos (\theta_{\rm QCD} )  + {\cal O}\left( y^{2} \right) ~. 
 \label{derivative of thetaQCD}
}
Therefore, the quantity $\cal Y$ in Eq.~(\ref{scalar amplitude}) is computed as
\aln{ {\cal Y}(\bar{\theta}_{\rm QCD})= 
\left. \left( \frac{\bar{\theta}_{\rm ini} }{\bar{\theta}_{\rm QCD} } \right)  
\middle/  \ 
 \overline{ \left( \frac{\delta \theta_{\rm ini} }{\delta \theta_{\rm QCD}}  \right)}
 \right.
= 1 +y \left(  \frac{\sin (\bar{\theta}_{\rm QCD})}{\bar{\theta}_{\rm QCD}} - \cos (\bar{\theta}_{\rm QCD}) \right)  +{\cal O}(y^{2}) ~. \label{thermal inflation Y}
}
These results show that the axion field does not evolve much until $T=T_{\rm QCD}^{}$. 

\vspace{5mm}
In terms of the averaged angle $\overline{\theta}_{\rm QCD}$ at the QCD scale, the ratio of the energy densities is given by
\aln{
R&=\frac{\rho_A^{}}{\rho_r^{}}\bigg|_{T=T_{\rm QCD}^{}}\simeq  \frac{30}{\pi^2 g_{\rm QCD}^{}}\frac{m_u^{}/m_d^{}}{(1+m_u^{}/m_d^{})^2}\frac{m_\pi^2f_\pi^2}{T_{\rm QCD}^4}  
(1-\cos (\overline{\theta}_{\rm QCD}) )\nn \\
&\simeq 0.012\times \left(\frac{150{\rm MeV}}{T_{\rm QCD}^{}}\right)^4  
(1-\cos (\overline{\theta}_{\rm QCD}) ) .  
\label{R in thermal inflation}
}
%The angle $\overline{\theta}_{\rm QCD}$ is related to $\overline{\theta}_{\rm ini}$ as seen above. 
Compared to the conventional case of Eq.~(\ref{R in standard QCD axion}), there is no small numerical factor since the axion field evolution is slow in the thermal inflation with a larger Hubble parameter, $3H_{\rm TI}>m_{A}$.  
Namely,  due to the smallness of $\eta$, the angle $\bar\theta$ does not so much decrease before the QCD transition.  
As a result, $R\sim 0.01$ is naturally realizable in the thermal inflation scenario with e.g., $T_{\rm TI} \sim $ TeV
as far as $\bar\theta_{\rm QCD}$ is around $\pi/2$.
Thus, one of the difficulties in the standard scenario is evaded. 

By substituting this $R$ and Eqs.~(\ref{anharmonicity X})(\ref{anharmonicity Y})(\ref{derivative of thetaQCD}) into Eq.~(\ref{CMB by axion}), 
the scalar spectrum amplitude of the CMB in Eq.(\ref{CMB by axion}) becomes
\aln{
\sqrt{A_s} \simeq  0.006\times \sin (\bar{\theta}_{\rm QCD} ) \left(\frac{150{\rm MeV}}{T_{\rm QCD}} \right)^4\frac{\overline{H}_\text{exit}^{}(k_*^{})}{4\pi f_A^{}}  \times \left( 1 - y \cos (\bar{\theta}_{\rm QCD})
 \right) 
 \label{AsTI}
} 
up to  ${\cal O}(y^2)={\cal O}( (\eta /8b \alpha)^2)$ contributions.
The axion scenario for the CMB fluctuation is realized if this gives the observational value $A_s \simeq 2.1 \times 10^{-9}$. 
From the non-Gaussianity constraint, $\bar\theta_{\rm QCD}$ must be around $\pi/2$. Thus the requirement
for $A_s$ gives a relation between $\overline{H}_{\rm exit}$ and $f_A$. 

%%%%%%%%%%%%%%%%%%%%%%%%%%%%%%%%%%%%%%%
\vspace{5mm}
\noindent$\bullet$ {\bf Corrections to non-Gaussianity}\\
The nonlinear evolution before the QCD slice also affects the non-Gaussianity. 
Solving Eq.~(\ref{ini-QCD}) with respect to $\theta_{\rm QCD}$, one finds $\theta_{\rm QCD} = \theta_{\rm ini} -y \sin(\theta_{\rm ini})$ up to ${\cal O}(y^{2})$ terms.
With this relation,
$\delta A_{\rm QCD}$'s in Eq.~(\ref{expansion non gaussianity}) is expanded with respect to the initial Gaussian fluctuation $\delta A_{\rm ini}$.
Then, for a small $y \ne 0$, the nonlinearity parameters become
\aln{
f_{\rm NL}(\bar{\theta}_{\rm QCD} , y) &= f_{\rm NL}(\bar{\theta}_{\rm QCD} , 0) - \frac{5}{3} F(\bar{\theta}_{\rm QCD} , y) ~, \label{f(yne0)} \\
g_{\rm NL}(\bar{\theta}_{\rm QCD} , y) &= g_{\rm NL}(\bar{\theta}_{\rm QCD} , 0) +\frac{25}{9} G(\bar{\theta}_{\rm QCD} , y) - \frac{10}{3} f_{\rm NL}(\bar{\theta}_{\rm QCD} , 0)   F(\bar{\theta}_{\rm QCD} , y) ~, \label{g(yne0)}
 }
where $f_{\rm NL}(\bar{\theta}_{\rm QCD} , 0)$ and $g_{\rm NL}(\bar{\theta}_{\rm QCD} , 0)$ 
are given by Eq.~(\ref{f(y=0)}) and Eq.~(\ref{g(y=0)}) respectively, and
\aln{
F(\bar{\theta}_{\rm QCD} , y) := 
 \frac{2}{f_{A} R} 
\left.\frac{\bar{V}}{\bar{V}'}\right|_{\rm QCD} 
\overline{ \left( \frac{\delta^{2} \theta_{\rm QCD}}{\delta \theta_{\rm ini}^{2}} \right)} \ 
 \overline{ \left(  \frac{\delta \theta_{\rm ini}}{\delta \theta_{\rm QCD}}\right)}^{2} 
 = \frac{2 y}{ R} \left( 1- \cos (\bar{\theta}_{\rm QCD}) \right) + {\cal O}(y^{2}) ~, \label{FG} 
 }
\aln{
G(\bar{\theta}_{\rm QCD} , y) &:= 
\frac{8}{3f_{A}^{2} R^{2}} \left.\frac{\bar{V}^{2}}{\bar{V}'^{2}}\right|_{\rm QCD}   
\overline{ \left( \frac{\delta^{3} \theta_{\rm QCD}}{\delta \theta_{\rm ini}^{3}} \right) } \
\overline{\left( \frac{\delta \theta_{\rm ini}}{\delta \theta_{\rm QCD}}  \right)}^{3} =  \frac{8 y}{3 R^{2}} \tan^{2} (\bar{\theta}_{\rm QCD} /2 ) \cos (\bar{\theta}_{\rm QCD}  )  + {\cal O}(y^{2}) ~.  
\nn
}
As one can easily see from these expressions, 
non-linear effects shift the positions of zeros of the function
$f_{\rm NL}(\bar{\theta}_{\rm QCD},y)$ from $|\theta_{\rm QCD}^{}|=\pi/2,\ 3\pi/2$. 
At the leading order of $y$ and $\delta \bar{\theta}_{\rm QCD}^{}=\bar{\theta}_{\rm QCD}^{}-\pi/2$, 
the position of its zero is given by  $|\bar{\theta}_{\rm QCD}| = \pi /2 + R + y + {\cal O}(y^{2})$.  
In general, we denote such a vanishing point as $\overline{\theta}_S^{}$. 
Suppose that future observations determine
the nonlinearity parameter as $f_{\rm NL}(\bar\theta_{\rm QCD},y)=f_{\rm NL}^{\rm obs.}$. 
Then, by solving this equation, the angle $\bar{\theta}_{\rm QCD}$ is determined in terms of $y$ and $R$ as
 $|\bar{\theta}_{\rm QCD}| \simeq |\bar{\theta}_{\rm ini}|-y  \simeq \pi/2 + y + R(1+ 3 f_{\rm NL}^{\rm obs.}/10)$
 up to ${\cal O}(y^{2})$ corrections.

%%%%%%%%%%%%%%%%%%
\vspace{5mm}
\noindent$\bullet$ {\bf Estimation of $r_A$}\\
Now let us  evaluate the present axion abundance $r_A$ by solving the evolution of the axion field after the QCD transition 
during the thermal inflation until $T_{\rm end}$ in Fig.\ref{fig:history}.  
The subsequent calculation after the thermal inflation is the same as in the standard scenario without thermal inflation:
the angle at the end of the thermal inflation $\theta_{\rm end}$ provides the ``initial condition'' for the standard thermal history of the axion field after the reheating. 

After temperature drops down to $T_{\rm QCD}^{}$ during the thermal inflation, 
the axion potential does not increase any more, 
and instead of Eq. (\ref{SR eom of axion 1}), 
the attractor EOM of axion is given by 
\aln{\frac{d\theta}{d\Delta } = - \eta\sin\theta  \ .  \label{SR eom of axion 2} }
with $\Delta := H_{\rm TI} \int_{t_{\rm QCD}}^{t}  \lambda dt = \log (T(t_{\rm QCD})/T) \geq 0$. 
This $\Delta$ takes a positive value since we are interested in $t \ge t_{\rm QCD}^{}$. 
Also note that after the QCD transition (i.e., after the axion potential is generated), 
we have an inhomogeneous temperature of $T(x, t_{\rm QCD}) \neq T_{\rm QCD}$ even at the QCD slice. 
It is because 
the energy transferred from radiation to the axion field  fluctuates inhomogeneously. 
%Temperature $T$ after the QCD slice at \red{$t=t_{\rm QCD}$ } is related to $\Delta$ as $T/T(t_{\rm QCD}) = e^{-\Delta }$. 
%Note that $T(t_{\rm QCD})$ is not uniform because  
%the  energy transfer 
%on the QCD slice makes the temperature fluctuate around the uniform $T_{\rm QCD}$.

We integrate this attractor equation  from $\Delta = 0$ to the ``end'' slice on which 
the temperature is given by a uniform value, $T=T_{\rm end}^{}$. 
Denoting the corresponding e-folding number by
\aln{\Delta_{\rm end} = \ln \frac{T(t_{\rm QCD}) }{ T_{\rm end}} \simeq 2.7+\log\left(\frac{T(t_{\rm QCD})}{150{\rm MeV}}\right)+\log\left( \frac{10{\rm MeV}}{T_{\rm end}^{}}\right)  ~, \label{Delta_end}}
we obtain  
\aln{
\theta_{\rm end} = 2\arctan\left[ \tan(\theta_{\rm QCD}/2)e^{-\eta \Delta_{\rm end} } \right]  \label{theta_end function of theta_QCD} 
~.}
It should be noted that, unlike Eq.~(\ref{thetaQCD}) before the QCD slice,
we have the interval of the integration $\Delta_{\rm end}$ in the above expression because the axion field gradually rolls down during the thermal inflation.  
Thus if the e-folding number after the QCD transition becomes larger, 
the axion field further rolls down  toward the minimum. 

Suppose that $\Delta_{\rm end}$ is sufficiently large so that we have a small value of $\theta_{\rm end}^{}$. 
This condition is required from Eq.~(\ref{theta_exit-r_A}) where the left-hand side identified as $\bar{\theta}_{\rm end}$ 
 and $r_{A}$  in the right hand side is very small as in Eq.(\ref{r and R}). 
Then, we have
\aln{\theta_{\rm end} \simeq 2 \tan(\theta_{\rm QCD}/2)e^{-\eta \Delta_{\rm end} } \label{theta_end}}
By differentiating it with respect to $\theta_{\rm QCD}^{}$, we get
\aln{\frac{\delta \theta_{\rm end}}{\delta \theta_{\rm QCD}} 
\simeq \frac{\theta_{\rm end} }{\sin (\theta_{\rm QCD})}  - \eta ~\theta_{\rm end} \frac{\delta \Delta_{\rm end}}{\delta \theta_{\rm QCD}}  \label{delta-theta_end}
}
Let us remember that, since the QCD/end slice is a uniform density/temperature slice and the primordial curvature perturbation $\zeta_{\rm inf} = -\psi_{\rm QCD}$ is assumed to be negligible, we have fluctuations of e-foldings $\delta N$ between these slices as
\aln{\delta N  =\Delta_{\rm end} - \bar{\Delta}_{\rm end} = \psi_{\rm end} = -\zeta_{r} ~, \label{''delta N''}}
where $\bar{\Delta}_{\rm end}$ is obtained by replacing $T(t_{\rm QCD})$ by the uniform temperature $T_{\rm QCD}$ in Eq.~(\ref{Delta_end}).
%\aln{\bar{\Delta}_{\rm end} \equiv \log\frac{T_{\rm QCD}^{}}{T_{\rm end}^{}}=2.7+\log\left(\frac{T_{\rm QCD}^{}}{150{\rm MeV}}\right)+\log\left(\frac{10{\rm MeV}}{T_{\rm end}^{}}\right). }
Since $\delta \theta_{\rm QCD} /\delta \theta_{\rm ini} \sim 1$, we find
\footnote{
%As mentioned around Eq.(\ref{''delta N''}) below, this can be understood as fluctuations of e-foldings of the thermal inflation. 
In this regard, the mechanism is similar to those discussed in \cite{Kawasaki:2009hp,Dimopoulos:2017qqn}. The crucial difference is that the temperature $T_{\rm end}$ at which the thermal inflation ends is assumed to fluctuate in \cite{Kawasaki:2009hp,Dimopoulos:2017qqn} as realizations of the ``end of inflation'' scenario \cite{Lyth:2005qk}; this is not the case here.
Note also that, generally speaking, QCD axion's fluctuations can cause those of $T_{\rm end}$ when the Higgs field, whose vacuum expectation value is responsible for axion's potential, is coupled with the flaton field. We simply assume such a contribution to fluctuations of e-foldings is negligibly small.
} 
 $\delta \Delta_{\rm end} /\delta \theta_{\rm QCD} \sim \sqrt{A_{s}} /(\overline{H}_{\rm exit}/2\pi f_{A}) \propto R$, and hence, the second term of Eq.~(\ref{delta-theta_end}) is negligible compared to the first one.

\vspace{5mm}
The angle $\theta_{\rm end} \ll 1$ gives the initial condition for the evolution after the thermal inflation. 
The evolution of the axion field before the oscillation can be neglected as in Section \ref{standard QCD axion}. 
%\red{
Since the axion potential is now well approximated by a harmonic potential and the evolution equation of the angle 
is almost linear\footnote{
The evolution is linear since the energy transfer between radiation and axion is negligible and the axion's energy density is small. 
See footnotes \ref{standard Y} and \ref{standard Z}.}, 
 the ratio $\delta \theta / \bar{\theta}$ does not change
 after the thermal inflation. 
Therefore, the quantity Eq.~(\ref{X_i}) is computed as
\aln{
{\cal Z}(\bar{\theta}_{\rm QCD})=\left( \frac{\bar{\theta}_{\rm QCD}}{\bar{\theta}|_{\rm late \ time}} \right)
\overline{ \left( \frac{\delta \theta|_{\rm late \ time}}{\delta \theta_{\rm QCD}} \right) }
\simeq \left( \frac{\overline{\theta}_{\rm QCD}}{\bar{\theta}_{\rm end}}\right)
 \overline{ \left(\frac{\delta \theta_{\rm end}^{}}{\delta \theta_{\rm QCD}}\right) }  
 \simeq   \frac{\bar{\theta}_{\rm QCD} }{\sin (\bar{\theta}_{\rm QCD}) } + {\cal O}(R) ~. \label{thermal inflation Z} 
}
In the calculation of the present axion abundance $r_A$, 
the same calculation  in Section \ref{standard QCD axion} can be  applied to the thermal inflation model
by replacing the initial angle $\bar\theta_{\rm ini}$ in Eq.~(\ref{theta_exit-r_A}) 
  with $\theta_{\rm end}$ in Eq.~(\ref{theta_end}). 
Putting $T_{\rm QCD}=150$MeV, 
we have the relation between $r_A$ and $\overline{\theta}_{\rm QCD}$;
\aln{
  r_A^{}&\sim \left(\frac{2\tan(\overline{\theta}_{\rm QCD}/2)}{0.3}\right)^2 \left(\frac{f_A^{}}{10^{12}\text{GeV}}\right)^{1.16} 
\exp\left(-2 \eta \bar{\Delta}_{\rm end}^{}\right) ~.
\label{axion abundance in thermal inflation} 
}
From the isocurvature constraint in Eq.(\ref{condition 2}), $r_A$ must be sufficiently small. This requires 
that the combination $\eta \bar\Delta_{\rm end} \simeq 30.2 \times y  \bar\Delta_{\rm end}$ must be sufficiently large.

\begin{figure}[t!]
\begin{center}
\includegraphics[width=8cm]{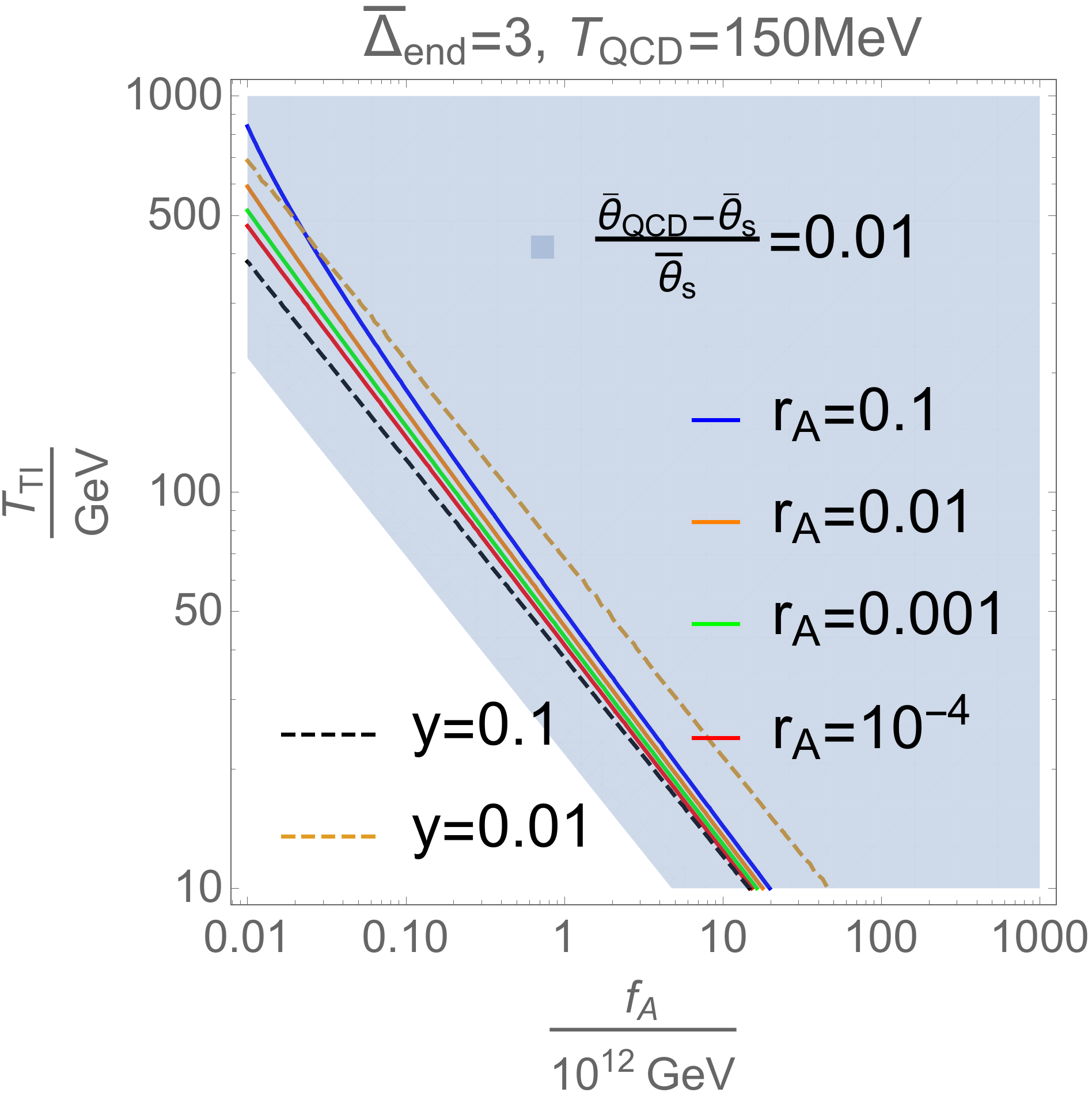}
\includegraphics[width=8cm]{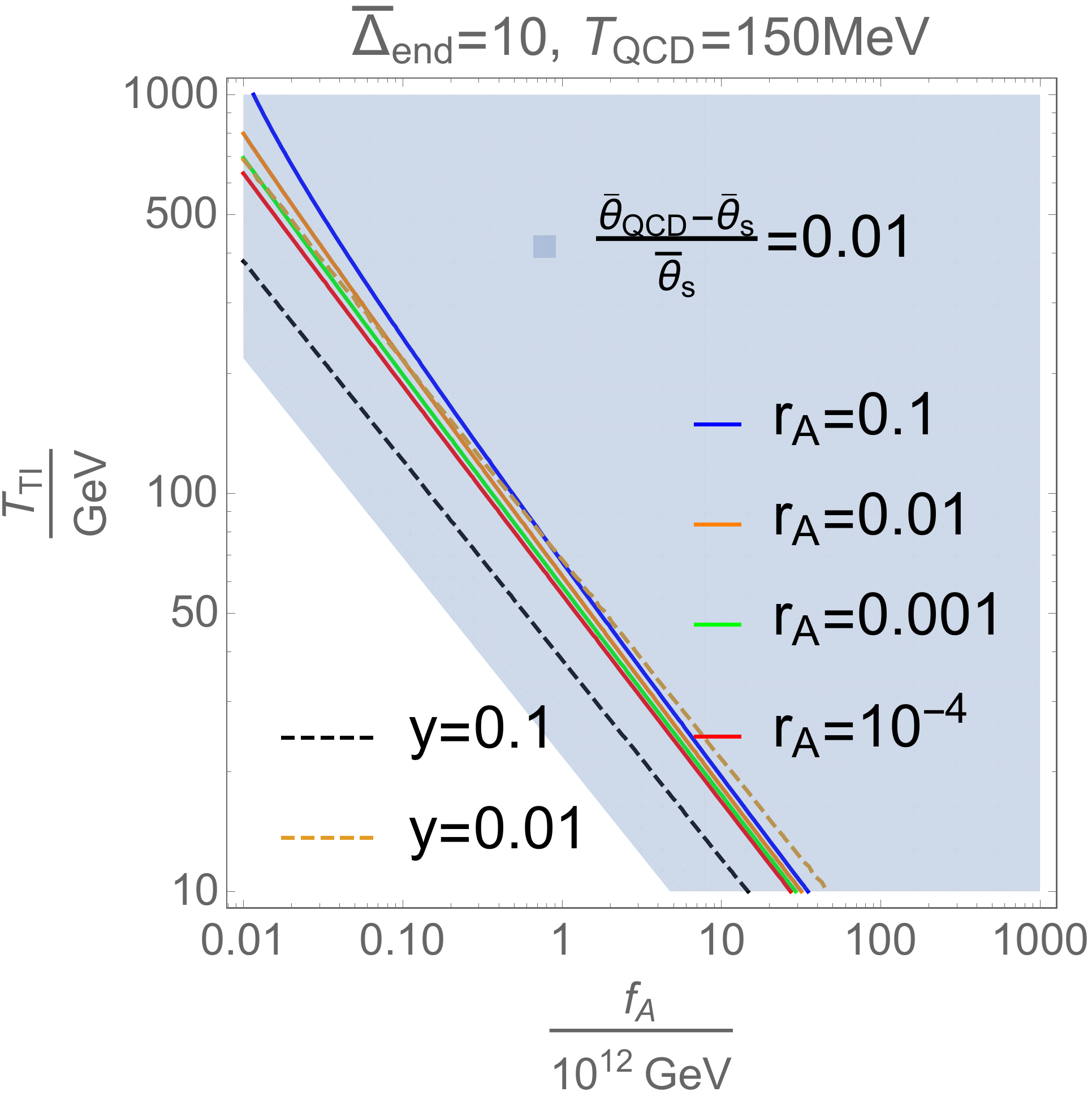}
\caption{%The CMB amplitude $A_s$ can be explained if parameters are on solid lines.
The CMB amplitude $A_s$ is explained if $\overline{H}_{\rm exit}$ is appropriately chosen as in Eq.(\ref{H_exit,TI}). 
On each solid line, a different value of $r_A$ in Eq.~(\ref{axion abundance in thermal inflation})
is predicted, which is constrained by the isocurvature constraint of Eq.(\ref{condition 2}). 
Other parameters are set as $(\overline{\theta}_{\rm QCD}^{}-\overline{\theta}_{S}^{})/\overline{\theta}_{S}^{}=0.01$, 
and $\overline{\Delta}_{\rm end}^{}=3\ (10)$ in the Left (Right) panel. 
If the parameters are in the blue regions,   the non-Gaussianity constraint Eq.~(\ref{f and g}) 
is satisfied. 
We also plot $y\simeq \eta/30.2$ 
to see why a small value of $r_A$ in Eq.~(\ref{axion abundance in thermal inflation})
is realizable. The dashed black (orange) line corresponds to $y=0.1\ (0.01)$. 
The figure shows that the three conditions in section \ref{parameter region} can be satisfied in the thermal inflation scenario.
}
\label{fig:TI}
\end{center}
\end{figure}

\vspace{10mm}
In Fig.\ref{fig:TI}, we plot the prediction of axion abundance $r_A$ as a function of two parameters $(T_{\rm TI}, f_A)$. 
%If parameters are on the solid lines, the scalar amplitude $A_s$ in Eq.(\ref{AsTI}) can explain the CMB anisotropy. 
The scalar amplitude $A_{s}$ in Eq.(\ref{AsTI}) explains the CMB anisotropy if $\overline{H}_{\rm exit}$ is chosen as
\aln{
\left( 1 - y \cos (\bar{\theta}_{\rm QCD})  \right)  \sin(\bar{\theta}_{\rm QCD} ) \times \frac{\overline{H}_\text{exit}^{}(k_*^{})}{\pi f_A^{}}  \simeq 0.03 \times  \left(\frac{T_{\rm QCD}}{150{\rm MeV}} \right)^4 ~. \label{H_exit,TI}
}
On each line, the model predicts a different value of $r_A^{}$, which is constrained by the condition of isocurvature
fluctuations in Eq.(\ref{condition 2}). 
In the left (right) panel, 
we set the e-folding of thermal inflation between the QCD phase transition until the end of the thermal inflation 
as $\overline{\Delta}_{\rm end}^{}=3\ (10)$. 
%the contours of $r_A^{}=const$ in the $(f_A^{},T_{\rm TI}^{})$ plane for a fixed value of $\overline{\Delta}_{\rm end}^{}$. 
%The left (right) panel corresponds to $\overline{\Delta}_{\rm end}^{}=3\ (10)$, 
We further set the angle $\bar\theta_{\rm QCD}$ to take 
$(\overline{\theta}_{\rm QCD}^{}-\overline{\theta}_{S}^{})/\overline{\theta}_{S}^{}=0.01$.
Then  the non-Gaussianity constraint Eq.~(\ref{f and g}) is satisfied if the parameters are in the blue region. 
Any tiny value of $r_A$ can be obtained by tuning the parameters on the solid lines.
This is because $y \simeq \eta/30.2$ can take a larger value by choosing the parameters,
as plotted by the dashed lines; we represent the $y \simeq \eta/30.2=0.1\ (0.01)$ contours by dashed black (orange) lines.   
Thus all of the three conditions in \ref{parameter region} can be satisfied. 
The difficulties of the standard QCD axion scenario are evaded because
the axion never  starts to oscillate during the thermal inflation 
because the condition $3H_{\rm TI}^{}>m_A^{}(T)$ is always satisfied in this parameter region. 
Thus, the thermal inflation scenario can meet the necessary conditions discussed in Section \ref{axion scenario} 
provided  the inflation lasts long enough below the QCD phase transition.

\vspace{5mm}
Finally we comment on the non-Gaussianity of isocurvature fluctuations. 
One can  evaluate isocurvature contributions to the non-Gaussianity of the gravitational potential introduced in \ref{Curvature perturbations and CMB observables} by using the evolution of $\theta$ in Eq.(\ref{theta_end function of theta_QCD}) between the QCD and end slices, as well as the one after the thermal inflation. 
  As discussed in \ref{app:isocurvatureNG} (in particular in Eqs.~(\ref{bound on W})(\ref{relation of f})), the effect 
  turns out to be very small since the isocurvature perturbation is tiny 
  compared to the adiabatic one. Thus the observational constraints for isocurvature non-Gaussianity 
  can be easily evaded.

%________________________________________________________
\subsection{$B$-$L$ model with QCD axion}\label{B-L}
In the final section, we briefly comment on a classically conformal  $B$-$L$ model 
%\cite{Iso:2009ss}-\cite{Iso:2017uuu} 
\cite{Iso:2009ss,Iso:2009nw,Iso:2012jn,Iso:2017uuu} 
with QCD axion because it has a potential to avoid difficulties in the standard QCD. 
A novel feature of the model is that, 
due to the assumption of the classical conformality (i.e. absence of quadratic terms in the scalar potential), 
the universe has experienced an era of supercooling for both of the $B$-$L$  and the electroweak 
symmetries \cite{Iso:2017uuu}:
these symmetries are not spontaneously broken 
until the chiral symmetry breaking $\langle \bar\psi \psi \rangle \neq 0$
occurs at the QCD temperature. Also the vacuum energy of a scalar field with a $B$-$L$ charge 
generates thermal inflation from TeV to QCD scales.
Once $\langle \bar\psi \psi \rangle$ acquires a non-zero value, 
a linear term in the Higgs potential is generated through the Yukawa coupling $y h \bar{\psi} \psi$. 
Thus, the electroweak symmetry breaking occurs at $T_{\rm QCD}$,  
but with a  smaller vev $\langle h \rangle \sim {\cal O}(\Lambda_{\rm QCD})$ than the electroweak scale. 
As the temperature of the universe further goes down, 
$B$-$L$ symmetry is then spontaneously broken 
by a scalar mixing of the SM Higgs $h$ and $B$-$L$ scalar field $\phi_{\rm B\text{-}L}$, and then
the SM Higgs acquires  an ordinary vev at $\langle h \rangle= 246$ GeV.  
Due to the vacuum energy $V(\phi_{\rm B\text{-}L})$ of the $B$-$L$ scalar $\phi_{\rm B\text{-}L}$, 
thermal inflation continues until the end of the supercooling era, which is 
below $T_{\rm QCD}$. 
Therefore, the scenario investigated in the previous section is naturally realized: 
axions are diluted in the thermal inflation to have a small value of $r_A$.  
This model provides 
 a natural scenario for the thermal inflation below the QCD scale, 
 and the classically conformal $B$-$L$ model with the QCD-like axion can be a good candidate
  for the scenario to explain CMB anisotropy by the axion isocurvature fluctuations. 
But there is one technical difficulty to obtain a large value of $R$ 
since quark masses are smaller than usual when the axion potential is generated, 
due to the smallness of $\langle h \rangle$ at $T=T_{\rm QCD}^{}$, 
and the axion energy density becomes smaller than the scenario in the previous section. 
In order to overcome this difficulty in generating sufficient amount of fluctuations, 
we need to raise the axion potential, e.g., by introducing a mixing of the SM Higgs and axions. 
Further studies of modifications of the model are left for future publications.

\section{Conclusions}
In this paper, we have investigated  a possibility to explain the CMB anisotropy by the primordial
fluctuations of QCD-like axions.  Such a scenario is generally characterized by three parameters, 
amplitude of the axion fluctuation $\overline{H}_{\rm exit}/f_A \bar\theta_{\rm ini}$, ratio of energy densities of axion to radiation 
$R=\Omega_A^{}/\Omega_r^{}%|_{T_{\rm QCD}}
$ when the axion potential is generated, and the fraction of the present 
axion abundance $r_A=\Omega_A/\Omega_{\rm CDM}|_{\rm today}$.
In order to be consistent with the CMB observations, especially the non-Gaussianity and isocurvature constraints,
 we need $0.01 \lesssim R  \le 1$ and simultaneously 
 a  small value of $r_A$. It is summarized in Fig. \ref{fig:region}.
To realize these values, a certain dilution mechanism of axions after $T_{\rm QCD}$ is inevitable, and
a natural possibility is the thermal inflation that lasts below the QCD scale. 
Specifying the thermal inflation model by its Hubble parameter $H_{\rm TI}$ (or its corresponding temperature $T_{\rm TI}$) and  the number of e-folding $\overline{\Delta}_{\rm end}$ from $T_{\rm QCD}$ to the end of the thermal inflation, 
we obtain the allowed parameter region of the thermal inflation model shown in Fig.\ref{fig:TI}. 
In this investigations, we have taken important effects of non-Gaussianity from the axion potential itself
and from the evolutions of the axion field before and after the QCD phase transition, 
 denoted respectively by ${\cal X}, {\cal Y}$ and ${\cal Z}$. 
In order to evade the non-Gaussianity constraint for $f_{\rm NL}$, the axion angle at the QCD scale must be $\overline{\theta}_{\rm QCD} \simeq \pi/2$. 
It is interesting  then that the non-Gaussianity $g_{\rm NL}^{}$ 
 becomes within reach of observations in near-future. 

 An interesting prediction of the axion scenario for the CMB anisotropy is that, as noted in the footnote 5, since the axion field did not have potential  when its fluctuation is generated during the primordial inflation, the spectral index $n_s^{} = 1-2 \epsilon$ coincides with the tensor mode index $n_t^{}$. 
Thus, when the tensor mode is discovered and its spectral index
is observed, we can justify/falsify our model. 

As a concrete model of a such thermal inflation scenario, we comment on a classically conformal $B$-$L$ model. 
This model naturally realizes thermal inflation at very low energy scale and can be a good candidate.
However, there is a technical difficulty to obtain sufficient amount of the CMB fluctuations. 
To overcome this difficulty, we need to raise the height of the axion potential, and accordingly the amplitude
of the Higgs vev  when the QCD phase transition occurs than $ \langle v \rangle \sim {\cal O}(\Lambda_{\rm QCD} )$. 
It may be possible by generating a negative thermal mass term of the Higgs from interactions with other
scalar fields.  
We want to come back to this model in future investigations.

%%%%%%%%%%%%%%%%%%%% ACKNOWLEDGMENTS %%%%%%%%%%%%%%%%%%%%
\section*{Acknowledgements} 
This work is supported in part by Grants-in-Aid for Scientific Research No. 16K05329 and No. 18H03708 from the Japan Society for the Promotion of Science. 
K.S. is supported by Grants-in-Aid for Scientific Research No. 16H06490. 
K.K. is supported by the Grant-in-Aid for JSPS Research Fellow, Grant Number 17J03848. 
%%%%%%%%%%%%%%%%%%%%%%%%%%%Appendix%%%%%%%%%%%%%%%%%%%%%%%%%%%%%%%%%%%%%%%%
\appendix 
\def\thesection{Appendix \Alph{section}}

%__________________________________________________________
\section{Curvature perturbations and CMB observables \label{Curvature perturbations and CMB observables}}
%________________________________________________________
%\subsection{CMB fluctuations and isocurvature perturbation}
In this appendix, we 
discuss how the gauge invariant curvature perturbations
 are related to the CMB observables in $\Lambda$CDM model. 
In the following, we consider only linear perturbations for simplicity.  
Assume there are two CDM components; one ($d$) is produced from the SM radiation in the early universe and the other 
($A$ for axion) is essentially decoupled from the SM in late time.
After the Big-Bang Nucleosynthesis, each of these CDMs, 
the SM baryon ($b$) and photon ($\gamma$) are described as a perturbed perfect fluid with 
\be
w_{\gamma}^{} = \frac{1}{3} ~~,~~~ w_{b} = w_{d} = w_{A} = 0 ~,
\ee
\be
\zeta_{r}^{(1)} := \zeta_{\gamma}^{(1)}=\zeta_{b}^{(1)}=\zeta_{d}^{(1)}~~,~~~  \zeta_{A}^{(1)} \ne  \zeta_{r}^{(1)} ~.
\ee
The matter dominated era starts at $T=T_{\rm eq}\sim {\rm eV}$ and the last scattering surface is given
at $T=T_{\rm lss}\sim 0.1$eV. 
The dark energy fraction $\Omega_{\Lambda}^{}$ is still totally negligible and the curvature perturbation Eq. (\ref{linear-curvature-perturbation-perfect-fluids}) is given by 
\be
\zeta^{(1)} = \frac{(4\Omega_{\gamma} + 3\Omega_{b} + 3 \Omega_{d})\zeta_{\gamma}^{(1)} + 3\Omega_{A} \zeta_{A}^{(1)} }{4\Omega_{\gamma} + 3\Omega_{b} + 3\Omega_{d}^{} + 3\Omega_{A}^{} }\bigg|_{\text{lss}}^{} 
=  \zeta_{\gamma}^{(1)} - \zeta_{\gamma,{\rm iso}}^{(1)} \label{LSS-linear-curvature-perturbation}
\ee
where
\be
\zeta_{\gamma,{\rm iso}}^{(1)} =&  \frac{\Omega_{A}^{} }{3\Omega_m^{}+ \Omega_{\gamma}}\bigg|_{\text{lss}}^{} S_{A}\simeq \frac{\Omega_A^{}}{3\Omega_m^{}}\bigg|_{\text{lss}}^{}S_{\rm A}^{}=\frac{\Omega_A^{}}{3\Omega_m^{}}\bigg|_{\text{today}}^{}S_{\rm A}^{},~~S_{\rm A}^{} := - 3(\zeta_{\rm A}^{(1)} -\zeta_{\gamma}^{(1)} ) ~. 
\ee 
Here, $\Omega_{m} := \Omega_{b} + \Omega_{d}+ \Omega_{A} = \Omega_{\rm CDM}+ \Omega_{A}$ is the total matter energy density fraction and the ratio $\Omega_A^{}/\Omega_m^{}$ can be evaluated at the present time  because it does not change much 
through the time evolution of the universe.\footnote{
Precisely speaking, $\Omega_{\gamma}$ is not negligible and its time evolution needs to be taken into account for further
systematic analysis.}
If the integrated Sachs-Wolfe contribution is neglected, the CMB temperature anisotropy is given by a simple analytic expression as \cite{Gordon:2002gv,Langlois:1999dw}
\be
\frac{\delta T}{T}  = -\zeta_{\gamma}^{(1)} - 2 \Phi^{(1)}\bigg|_{\rm lss}^{} \label{T-anisotropy}
\ee
with the gravitational potential $\Phi$ which is, in the matter dominated era, approximated as
\be
\Phi \simeq  -\frac{3}{5} \zeta  ~. \label{Weyl potential}
\ee
Plugging this and Eq. (\ref{LSS-linear-curvature-perturbation}) into Eq. (\ref{T-anisotropy}), we find 
\be
\frac{\delta T}{T} = -\zeta_{\gamma}^{(1)} +\frac{6}{5} \zeta^{(1)}\bigg|_{\rm lss} = \frac{1}{5} \zeta_{\gamma }^{(1)} - \frac{6}{5} \zeta_{\gamma,{\rm iso}}^{(1)}\bigg|_{\rm lss}  ~,
\ee
which shows that the isocurvature component %in photon's perturbation 
affects the CMB anisotropy via the gravitational potential at the last scattering.
With the definition of ${\cal R}$ and ${\cal I}$ given in (\ref{definition of R and I}),
the observation of the two point function of the temperature perturbation gives the constrains (\ref{Planck-scalar}), (\ref{uncorrelated}) and (\ref{correlated}).

%\mage{More on SW effect...}

\

Furthermore, the Planck experiment gives constraints on the three point function of the gravitational potential as follows. 
Here, (\ref{Weyl potential}) is divided into to parts as $\Phi = \Phi^{\rm a}  + \Phi^{\rm i}$ where 
%\be
%\Phi = \Phi^{\rm a}  + r_{c} \Phi^{\rm i} ~,
%\ee
\be
\Phi^{a} := \frac{3}{5} {\cal R} ~~~, ~~~~ \Phi^{\rm i} :=  \frac{r_{c}}{5} {\cal I} \label{Phi^a Phi^i}
\ee
with $r_{c} := [ \Omega_{\rm CDM} + \Omega_{b} ]_{\rm today} \simeq 0.842$.
In general, both can be significant in the three point function \cite{Bartolo:2001cw,Kawasaki:2008sn,Langlois:2008vk,Kawasaki:2008pa,Hikage:2008sk}.
Each component is expanded as
\be
\Phi^{J} =  \Phi^{J}_{u} \delta \phi^{u} + \frac{\Phi^{J}_{uv}}{2} \delta \phi^{u} \delta \phi^{v} + \cdots 
\ee 
with the repeated indices $u,v$ summed over scalar (inflaton and axion) fields $\phi^{u} = (\sigma, A)$ which acquire the almost scale invariant spectrum during the primordial inflation.
The local type bispectrum $B^{IJK}$ is defined by
\be
\langle \Phi^{I}(k_{1}) \Phi^{J}(k_{2}) \Phi^{K}(k_{3}) \rangle = (2 \pi)^{3} \delta^{(3)}(k_{1}+k_{2}+k_{3}) B^{IJK}(k_{1},k_{2},k_{3}) ~,
\ee
\be
B^{IJK}(k_{1},k_{2},k_{3}) = 2 f^{I,JK}_{\rm NL} P_{\Phi}(k_{2}) P_{\Phi}(k_{3}) + 2 f^{J,KI}_{\rm NL} P_{\Phi}(k_{3}) P_{\Phi}(k_{1}) + 2 f^{K,IJ}_{\rm NL} P_{\Phi}(k_{1}) P_{\Phi}(k_{2}) ~,
\ee
normalized in terms of the power spectrum
\be
P_{\Phi}(k) \simeq  \frac{2 \pi^{2}}{k^{3}} \times \frac{H^{2}}{4\pi^{2}}  (\Phi^{\rm a}_{u} \Phi^{\rm a}_{u}) = \frac{2 \pi^{2}}{k^{3}} \times  \frac{9 A_{s}}{25} ~,
\ee
where the isocurvature mode is assumed to have a negligible contribution to the power spectrum to be consistent with the observation that $\beta_{\rm iso} \ll1$.
Then, the six independent non-linear parameters \cite{Langlois:2011hn,Langlois:2012tm} are %approximately\footnote{In practice, we can regard this as a definition of the non-linear parameters.}
given as
\be
f^{I,JK}_{\rm NL} = \frac{ \Phi^{I}_{uv} \Phi^{J}_{u}\Phi^{K}_{v}}{2 (\Phi^{\rm a}_{u'} \Phi^{\rm a}_{u'})^{2}} ~. \label{fNL}
\ee
%For the fully anti-correlated case with
%\be
%{\cal I} = -y \times  {\cal R} ~~~, ~~~~ y \equiv \frac{4 r_{A}}{R|_{T=T_{A}}} < 3.3 \times 10^{-2} ~, \label{y}
%\ee
%we have
%\be
%f^{\rm a,aa}_{\rm NL} = -\frac{3}{y} f^{\rm a,ai}_{\rm NL} = -\frac{3}{y} f^{\rm i,aa}_{\rm NL} = \left(\frac{3}{y}\right)^{2} f^{\rm a,ii}_{\rm NL} = \left(\frac{3}{y}\right)^{2} f^{\rm i,ia}_{\rm NL} = -\left(\frac{3}{y}\right)^{3} f^{\rm i,ii}_{\rm NL} ~.
%\ee
%Therefore, once the purely adiabatic one satisfies the constraint $f^{\rm a,aa}_{\rm NL} = 4 \pm 10$ by Planck 2018 \cite{Akrami:2019izv}, the others are safely small. 
When only one scalar field (axion $A$ in our case) contributes to the large scale perturbation, we find
\be
f^{\rm a,aa}_{\rm NL} =  \frac{5}{6} \frac{ {\cal R}_{AA}}{{\cal R}_{A}^{2}} ~,
\ee
which corresponds to the coefficient $f_{\rm NL}$ of the second order term in the expansion (\ref{expansion of zeta}).
%\be
%\zeta_{\gamma} = \zeta_{\gamma {\rm L}} - \frac{3}{5} f_{\rm NL} \zeta_{\gamma {\rm L}}^{2} + \frac{9}{25}  g_{\rm NL}   \zeta_{\gamma {\rm L}}^{3} + \cdots ~.
%\ee
%The minus sign opposite to the one in \cite{Sasaki:2006kq} comes from the difference in the definition of $\zeta_{\gamma}$.
From Table 7 in \cite{Akrami:2019izv}, we see that
the constraint on this purely adiabatic component is the most stringent and given in Eq.~(\ref{f and g}).

\

%%%%%%%%%%%%%%%%%%%%%%%%%%%%%%
%\section{Standard Scenario of QCD Axion}
\section{Gradual energy transfer taken into account}\label{app:gradual}
In the body of paper, it is assumed 
that the energy transfer between the radiation and the axion field occurs all at once on the transition slice characterized by $T=T_{A}$ for brevity.
Here, let us consider a more realistic process of gradual energy transfer focusing on the standard QCD axion scenario
with $T_{A} =T_{\rm QCD}$. We will find that the gradual energy transfer does not change our conclusion so much.
%
%When the Higgs field acquires its vev at $T=T_{\rm EW}$, tiny but finite axion's potential is generated. 
%Since then, the energy density is continuously transferred from radiation to axion.
%We still rely on the sudden transition approximation that some quantities change discontinuously, for instance, axion's equation of state $w_{A}=\bar{\rho}_{A}/\bar{P}_{A}$ suddenly changes from $-1$ to $0$ at $T=T_{\rm osc}$ when the condition $m_{A} \ge 3 H$ is satisfied and the oscillation starts.
%Then, the evolution (energy conservation) equations are easily integrated.

\begin{figure}[t!]
\begin{center}
\includegraphics[width=12cm]{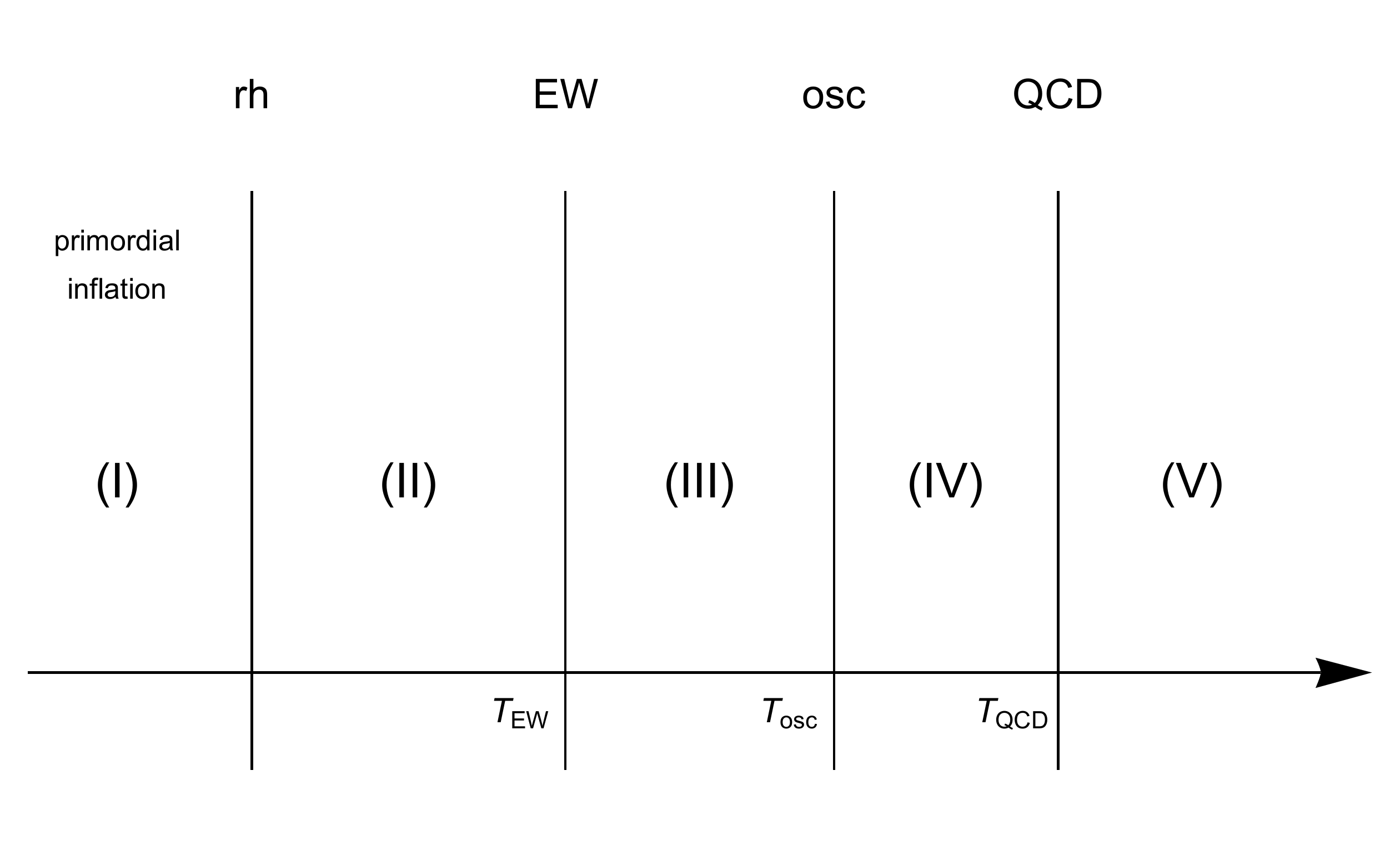}
\caption{Thermal history of standard QCD axion scenario.
The horizontal axis represents the direction of time evolution. 
}
\label{fig:Appendix}
\end{center}
\end{figure}

In the following, thermal history is divided into several stages by time slices on which the discontinuities take place.
For completeness, we start from the horizon exit of the CMB scale during the primordial inflation, stage (I).
Assuming the reheating is instantaneous, we have the radiation dominated era with (I\!I) $T>T_{\rm EW}$, (I\!I\!I) $T_{\rm EW} > T > T_{\rm osc}$, (I\!V) $T_{\rm osc} > T > T_{\rm QCD}$ and (V) $T<T_{\rm QCD}$.
%The energy densities of the radiation and the axion in a period (i) are denoted with superscript i as $\rho^{\rm i}_{r/A}$ and so are the associated curvature perturbations $\zeta^{\rm i}_{r/A}$; otherwise, their possible discontinuity due to the sudden transition approximation may cause confusions. \mage{読み直して、もし必要なさそうなら消します。}
And the four time slices connecting the stages are called, reheating (rh), electroweak (EW), oscillation (osc) and QCD slice, respectively, see Fig.\ref{fig:Appendix}. 
The total energy density $\rho$ and the spacetime metric, especially $\psi$, are continuous on each slices.
%and hence, there is no need to put such a superscript i.

For the standard QCD axion scenario, we are particularly interested in $\rho_{r}$ and $S_{A} =-3 (\rho_{A} -\rho_{r})$ in the last stage (V).
The former is nothing but $-\psi$ on a uniform temperature slice by definition and the latter is given as the perturbation of $\rho_{A}$ on a uniform temperature slice after the QCD transition, as discussed around (\ref{late time S_A}).

\

\noindent (I) Primordial inflation

The total energy is stored in the inflaton field $\sigma$ in this stage: $\rho = \rho_{\sigma}$.
Inflaton's equation of motion is $(-g)^{-1/2} \partial_{\mu} \left\{ (-g)^{1/2} g^{\mu\nu} \partial_{\nu} \sigma \right\} = \partial_{\sigma} V_{\sigma}$ with $V_{\sigma}$ being inflaton's potential.
%\be
%{\cal N}^{-2} \left[ \ddot{\sigma} + (3 {\cal N} H - \dot{\cal N}/{\cal N}) \dot{\sigma} \right] + V_{\sigma}' = 0 ~,
%\ee
%where we have left the lapse function ${\cal{N}}$ in general.  
%
Dropping the second time derivative, we get the slow-roll equation in the metric (\ref{FLRW}) as
\aln{3 H \dot{\sigma} /\lambda \simeq - \partial_{\sigma}  V_{\sigma}}
where $H^{2} \simeq V_{\sigma}/3 m_{\rm Pl}^{2}$ %with $m_{\rm Pl}$ the reduced Planck mass
and the dot denotes $t$-derivative. 
Since the energy density and the pressure density are given as $\rho_{\sigma} \simeq (\dot{\sigma} / \lambda )^{2} /2 + V_{\sigma}$ and $P_{\sigma} \simeq (\dot{\sigma} / \lambda )^{2} /2 - V_{\sigma}$ in the superhorizon limit,
the EOS is now approximated as $w_{\sigma} \simeq -1 + 2 \epsilon/3$ where $\epsilon := \left( m_{\rm Pl}  \partial_{\sigma} \ln V_{\sigma} \right)^{2} /2$ is the first slow-roll parameter. %and the inflationary expansion lasts as long as $\epsilon \leq 1$.

On the initial ``exit'' slice, $\psi_{\rm exit} = 0 \label{on-exit-slice}$ is assumed.
Following the definition (\ref{zeta_X}), we obtain
\aln{\zeta_{\rm inf} \simeq - \psi_{\rm exit} - \int^{\sigma_{\rm exit}}_{\bar{\sigma}_{\rm exit}}  \frac{d\sigma V_{\sigma}'}{ 2 \epsilon V_{\sigma}} = - {\rm sign}(V_{\sigma}') \int^{\sigma_{\rm exit}}_{\bar{\sigma}_{\rm exit}}  \frac{d\sigma / m_{\rm Pl} }{ \sqrt{2 \epsilon }} \label{zeta_inf}}
which conserves until the end of the inflation characterized by $\epsilon =1$.
Therefore, this ``reheating'' slice is nothing but a uniform $\rho_{\sigma}$ slice on which \aln{\psi_{\rm rh} = - \zeta_{\rm inf}~. \label{psi_rh}}
%At linear order, this becomes $\zeta_\sigma^{(1)}=-{\rm sign}(V_{\sigma}')  \delta \sigma/ ( m_{\rm Pl}^{} \sqrt{2 \epsilon })$ and we have 
%\aln{\langle \zeta_{\sigma \mathbf{k}}^{(1)}\zeta_{\sigma \mathbf{q}}^{(1)}\rangle =\frac{\langle \delta \sigma_{\mathbf{k}}^{(1)}\delta \sigma_{\mathbf{q}}^{(1)}\rangle}{2M_{pl}^2\epsilon}=(2\pi)^3\delta^{(3)}(\mathbf{k}+\mathbf{q})\times \frac{H^2}{4M_{pl}^2\epsilon k^3}=(2\pi)^3\delta^{(3)}(\mathbf{k}+\mathbf{q})\times \frac{V}{12M_{pl}^4\epsilon k^3}
%}
%which correctly reproduces  Eq. (\ref{adiabatic from inflation}).  
%
%As discussed above, this quantity conserves until the end of inflation where $\epsilon = 1$. 
%
%Let us call it ``reheating'' slice on which 
%\be
%\delta \rho |_{\rm rh}= \delta \rho^{\rm I}_{\sigma} |_{\rm rh} = 0 ~, \label{on-reheating-slice-I}
%\ee
%corresponding to $\delta \sigma = 0$, should hold.
%Therefore, we get
%\be
%\psi_{\rm rh} = -\zeta_{\sigma}^{\rm I} ~.
%\ee

During this stage, axion's primordial large scale fluctuations (\ref{gaussianA}) are also generated.
However, it is assumed that its potential is still absent and the field value does not evolve.

\noindent (I\!I) $T>T_{\rm EW}$ radiation domination

On the reheating slice, it is assumed that all the energy density is converted to radiation's energy density: $\rho= \rho_{r}$, and then, the curvature perturbation (\ref{zeta_inf}) is copied, via (\ref{psi_rh}), into the radiation:  \aln{\zeta_{r} = - \psi_{\rm rh}}
which is constant throughout this stage ending at $T=T_{\rm EW}$.
On this ``EW'' slice, the temperature is constant:
\be
\delta \rho |_{\rm EW}= \delta \rho_{r} |_{\rm EW} = 0 ~, \label{on-EW-slice-II}
\ee
and hence, we have
\be
\psi_{\rm EW} =  -\zeta_{r} =\psi_{\rm rh}  ~. \label{psi_EW}
\ee
%where $\zeta_{\rm inf}$ is the curvature perturbation originated from the primordial inflaton field, assumed to be negligibly small.

Regarding the axion field, its potential is still absent and the field value does not evolve with time.
And the perturbation $\delta A$ does not depend on the choice of time slice.

\ 

%%%%%%%
\noindent (I\!I\!I) $T_{\rm EW}>T>T_{\rm osc}$ % radiation domination 

The axion acquires the temperature dependent potential (\ref{QCD axion potential}) which is, however, too small to drive the field value to roll down to the minimum.
We simply assume $\dot{A} = 0$, and thus, $w_{A}=-1$ in this stage.
Then, $\rho_{A} = V_{A}$, we have
\be
\dot{\rho}_{A}  =  + \dot{T} \partial_{T} V_{A} = +8b \dot{\tau} V_{A} ~, \label{rho_A evolution III}
\ee
where $\tau := -\ln T/T_{\rm ref}$ with an arbitrary reference temperature $T_{\rm ref}$.
By the energy conservation, 
\be
\dot{\rho}_{r} + 4\lambda H \rho_{r} = - 8b \dot{\tau} V_{A}  \label{rho_r evolution III}
\ee
holds. We now solve it with respect to $\tau$.
With the local $e$-folding number $N=\int dt \lambda H$,
the above equation is rewritten as
\be
dN = d \tau \left[ 1- 2b  x_{0} e^{(8 b + 4)\tau} \right] ~,
\ee
where $x  := \rho_{A}/ \rho_{r} \propto e^{(8b +4)\tau}$ and $x_{0}  = x |_{\tau = 0} $.
Integrating it from the EW slice to the ``oscillation'' slice where $m_{A} /3 H = 1$, we have
\be
N_{\rm osc} - N_{\rm EW} &=& \int^{\tau_{\rm osc}}_{\tau_{\rm EW}} d \tau \left[ 1- 2b  x_{0} e^{(8 b + 4)\tau} \right] \label{N_osc-N_EW} \\
&=&  \tau_{\rm osc} - \tau_{\rm EW}  - \frac{b/4}{b +1/2} \left[ x_{\rm osc} -  x_{\rm EW} \right] ~. \notag
\ee
If tiny values of $x$ are neglected, this is nothing but $T \propto a^{-1}$ and $x \propto a^{8b+4}$.

The boundary condition on the EW slice (\ref{on-EW-slice-II}) becomes, in this stage,
\be
\delta \rho |_{\rm EW}= \delta \rho_{r} |_{\rm EW} + \delta \rho_{A} |_{\rm EW}  = 0 ~, \label{on-EW-slice-III}
\ee
which is equivalent to
\be
\delta \tau_{\rm EW}  = \frac{x_{\rm EW}/4}{1-2b x_{\rm EW}} \left. \frac{\delta V_{A0}}{V_{A0}} \right|_{\rm exit}  ~.
\ee
On the other hand, on the oscillation slice, perturbations should satisfy $\delta (m_{A} /3 H) = 0$, which is to say,
\be
\left[ \frac{\delta \rho_{r} }{\rho_{r}} + \frac{x}{1 + 2b /z } \frac{\delta \rho_{A} }{\rho_{A}} \right]_{\rm osc} =0    \label{on-osc-slice-III}
\ee
with $z := \rho_{r}/\rho = (1+ x)^{-1} $.
%To the first order in $x^{\rm I\!I\!I} \ll 1$, 
This is equivalent to
\be
\delta \tau_{\rm osc} = \frac{z_{\rm osc} x_{\rm osc} /4 }{2b+ z_{\rm osc} (1-2b x_{\rm osc} ) }  \left. \frac{\delta V_{A0}}{\bar{V}_{A0}} \right|_{\rm exit} ~. \label{delta tau on osc slice}
\ee
By differentiating (\ref{N_osc-N_EW}), we obtain ``$\delta N$'' as, to linear order,
\be
\psi_{\rm osc}-\psi_{\rm EW} &=&  \left[ 1- 2b x_{\rm osc}  \right] \times \delta \tau_{\rm osc} - \left[ 1- 2b x_{\rm EW}  \right] \times \delta \tau_{\rm EW}  \notag \\
&&- \frac{b/4}{b +1/2} \left[x_{\rm osc} - x_{\rm EW} \right] \times  \left. \frac{\delta V_{A0}}{V_{A0}} \right|_{\rm exit} \notag \\
&\simeq&  -\frac{b-1/2}{b +1/2} \frac{\bar{x}_{\rm osc}}{4} \times 2 \left. \frac{\delta A}{\bar{A}} \right|_{\rm exit} ~. \label{deltaN-III}
\ee
%where $R_{\rm osc} \equiv \bar{x}_{\rm osc} =  \bar{\rho}_{A}/ \bar{\rho}_{r} |_{\rm osc}$.
On the last line, terms with $\bar{x}_{\rm EW} \ll \bar{x}_{\rm osc}$ are omitted. 

\

%%%%%%%%%%%%%%%%%%%%%%
\noindent (I\!V) $T_{\rm osc}>T>T_{\rm QCD}$ %radiation domination 

In this stage, we assume the separation of two time scales; the frequency of axion's coherent oscillation is much larger than its changing rate.
Then, the time-averaging commutes with the partial $t$-derivative acting on the potential: $\langle \partial_{t} V_{A} \rangle_{t} = \partial_{t} \langle  V_{A} \rangle_{t} = \dot{T} \partial_{T} \langle  V_{A} \rangle_{t}$.
It is also assumed that $V_{A}$ is well approximated by a harmonic form, and then, $w_{A}=0$ and $ \rho_{A} = 2 \langle  V_{A} \rangle_{t}$.
Therefore, we have
\aln{
\dot{\rho}_{A}  + 3  \lambda H \rho_{A}  =  + \dot{T} \langle \partial_{T} V_{A} \rangle_{t} = + 4b \dot{\tau} _{A} \rho_{A} ~, \label{energy-transfer-IV-A}}
and from the energy conservation,
\be
\dot{\rho}_{r} + 4\lambda H \rho_{r} = - 4b \dot{\tau} _{A} \rho_{A}  ~. \label{energy-transfer-IV-r}
\ee
The second equation (\ref{energy-transfer-IV-r}) is written as
\be
d\tau / dN = (1-b x )^{-1} ~. \label{dtau/dN}
\ee
%where $x \equiv \rho_{A} / \rho_{r} $.
By eliminating $d \tau/dN$ from (\ref{energy-transfer-IV-A}),
we get
\be
dN = \frac{dx }{x } \frac{(1-bx )}{1+b(4+3x )} ~,
\ee
and combining this with (\ref{dtau/dN}),
\be
d\tau = \frac{dx}{x} \frac{1}{1+b(4+3x )} ~.
\ee
These equations are integrated from the oscillation slice to the QCD slice as
\be
N_{\rm QCD} - N_{\rm osc} = \frac{1}{1+4b} \left[ \ln \frac{x_{\rm QCD}}{x_{\rm osc}} - \frac{4(1+b)}{3} \ln \frac{1+b(4 + 3 x_{\rm QCD})}{1+b(4 + 3 x_{\rm osc})} \right] ~, \label{N_QCD-N_osc}
\ee
\be
\tau_{\rm QCD} - \tau_{\rm osc} = \frac{1}{1+4b} \left[ \ln \frac{x_{\rm QCD}}{x_{\rm osc}} - \ln \frac{1+b(4 + 3 x_{\rm QCD})}{1+b(4 + 3 x_{\rm osc})} \right] ~. \label{tau_QCD-tau_osc}
\ee
If tiny values of $x$ are set to zero in the second terms,
one finds $T  \propto a^{-1}$ and $\rho_{A} \propto a^{4b-3}$ which is consistent with (\ref{rho_A}).

On the oscillation slice, the sudden change of axion's EOS occurs.
Then, we have (\ref{delta tau on osc slice}), and equivalently,
\be
\frac{\delta x_{\rm osc} }{x_{\rm osc}}= \frac{1+x_{\rm osc} +2b/z_{\rm osc} }{x_{\rm osc} /4 } \times \delta \tau_{\rm osc} ~. %= \frac{ 2b + z_{\rm osc} (1+ x_{\rm osc} )  }{2b+ z_{\rm osc} (1-2b x_{\rm osc} ) }  \left. \frac{\delta V_{A0}}{\bar{V}_{A0}} \right|_{\rm exit}  ~. 
\ee
%\be
%\delta \tau_{\rm osc} \simeq \frac{R_{\rm osc}/8}{b+1/2}  \left. \frac{\delta V_{A0}}{\bar{V}_{A0}} \right|_{\rm exit}  ~~~,~~~~ \left. \frac{\delta R}{\bar{R}}\right|_{\rm osc} = (1+ R_{\rm osc})  \left. \frac{\delta V_{A0}}{\bar{V}_{A0}} \right|_{\rm exit}  ~. 
%\ee
On the QCD slice, the boundary condition is simply given by
\be
\delta \rho_{r} |_{\rm QCD} = 0 ~, \label{on-QCD-slice-IV}
\ee
therefore,
\be
\delta \tau_{\rm QCD} = 0 ~~~, ~~~~ \left. \frac{\delta x }{x }\right|_{\rm QCD} =   \left. \frac{\delta V_{A0}}{V_{A0}} \right|_{\rm QCD}  ~.  \label{delta-tau=0}
\ee
Differentiating (\ref{tau_QCD-tau_osc}) with respect to $\tau$ and $x$, we find
\be
\left. \frac{\delta V_{A0}}{V_{A0}} \right|_{\rm QCD} &=& \left\{ 1 + b(4+3 x_{\rm QCD}) \right\} \left[ \frac{\delta x / x }{1 + b(4+3 x )} - \delta \tau \right]_{\rm osc} \notag \\
&\simeq & \frac{ 1 + b(4+3 R )}{1 + 4b} \left. \frac{\delta V_{A0}}{\bar{V}_{A0}} \right|_{\rm exit} ~, \label{deltaV/V_QCD}
\ee
where $R :=  \bar{x}_{\rm QCD}   \ll 1$.
On the second line, only the first order of $\delta$ is retained, and then, $\bar{x}_{\rm osc} \ll R$ is neglected. 
And by differentiating (\ref{N_QCD-N_osc}), we obtain ``$\delta N$'' as, to linear order,
\be
\psi_{\rm QCD}-\psi_{\rm osc} &=& \left[ \frac{\delta x }{x } \frac{(1-b x )}{1+b(4+3 x )} \right]^{\rm QCD}_{\rm osc} \notag \\
&\simeq & - \frac{b R }{1 + 4b} \times 2 \left. \frac{\delta A}{\bar{A}} \right|_{\rm exit}  ~. \label{deltaN-IV}
\ee

\

%%%%%%%%%%%%%%%%%%%%%%
\noindent (V) $T_{\rm QCD}>T$ %radiation domination 

Here, we do not consider the vacuum energy to be released on the QCD phase transition.
Then, only the time derivative of axion's potential gets discontinuity.
Since it is the uniform temperature slice,
%the boundary condition should be
%\be
%\delta \rho_{r} |_{\rm QCD}= 0 ~,  ~~~,~~~~\delta \rho_{A}^{\rm V} |_{\rm QCD} = \delta \rho_{A}^{\rm I\!V}  |_{\rm QCD} ~. 
%\label{on-QCD-slice-V-2}
%\ee
we have $\delta \rho_{r} |_{\rm QCD}= 0$ and, to linear order,
\be
\zeta_{r} &=& - \psi_{\rm QCD} = - (\psi_{\rm QCD} - \psi_{\rm osc}) - (\psi_{\rm osc} - \psi_{\rm EW}) - \psi_{\rm EW} \notag \\
&\simeq & \zeta_{\rm inf} + \frac{2 b R}{1 + 4b}  \left. \frac{\delta A}{\bar{A}} \right|_{\rm exit} ~.
\ee
On the second line, the contribution (\ref{deltaN-III}) proportional to $\bar{x}_{\rm osc}$ is neglected.
Comparing this with (\ref{scalar amplitude}) for the harmonic case ${\cal X}_{\rm a} = 1$, we find the coefficient of $[\delta A/\bar{A}]_{\rm exit}$ now smaller by a factor $4b/(4b+1)$. The instantaneous energy transfer limit corresponds to $b \to \infty$, and then, this factor goes to unity as expected.

As for axion's fluctuation, since $w_{A} =0$ and there is no longer energy transfer in this stage, the evolution equation is linear and $\delta A/A$ conserves.
Then, from (\ref{deltaV/V_QCD}), we find
\aln{ \left. \frac{\delta V_{A0}}{V_{A0}} \right|_{\rm late~time} =  \left. \frac{\delta V_{A0}}{V_{A0}} \right|_{\rm QCD} \simeq  \frac{ 1 + b(4+3 R )}{1 + 4b} \left. \frac{\delta V_{A0}}{\bar{V}_{A0}} \right|_{\rm exit} ~. }
Note that the factor ${\cal X}_{\rm i} = 1 + 3R/4$ mentioned above (\ref{R in standard scenario}) is correctly reproduced in the instantaneous limit.
%%%%%%%%%%%%%%%%%%%%%

%%%%%%%%%%%%%%%%%%%%%%%%%
%______________________________
\section{Curvature perturbations in thermal inflation scenario}\label{final fluctuation}
In this appendix, we briefly discuss how the gradual energy transfer is taken into account in computing the curvature perturbation generated at the QCD transition during the thermal inflation, and how it is converted to the final one of radiation we observe today. 
We assume that all of the vacuum energy $V_{\rm TI} = (g_{\rm TI} \pi^{2}/30) T_{\rm TI}^{4}$ which drives the thermal inflation is suddenly converted to the energy density $\rho_{\phi}$ of a scalar field $\phi$ oscillating with its EOS $w_{\phi} = 0$.
%the equation of state of the inflaton $\phi$ suddenly changes from $-1$ to $0$ right after the end of thermal inflation,
And subsequently, this coherent oscillation suddenly decays to the radiation once the Hubble expansion rate drops to satisfy $3 H= \Gamma_{\phi}$ with $\Gamma_{\phi}$ being a constant decay rate of the scalar field: the gradual energy transfer from the oscillation to the radiation is neglected for simplicity since it is not our main focus here.

Let us apply the analysis in \ref{app:gradual} to the thermal inflation scenario where,
typically, the axion field does not oscillate around the potential minimum during the thermal inflation.
%$H\simeq H_{\rm TI} \equiv (V_{\rm TI}/3 m_{\rm Pl}^{2})^{1/2}$
Note that the considerations in \ref{app:gradual} are all valid regardless of the dominant component (apart from $z = 1-x$ introduced in (\ref{on-osc-slice-III}) which is now irrelevant). 
We can integrate the evolution equations (\ref{rho_A evolution III}) and (\ref{rho_r evolution III}) from the EW slice to the QCD slice on which $\delta \rho_{r}|_{\rm QCD} = 0$, hence, Eq.~(\ref{delta-tau=0}). 
Then, instead of Eq.~(\ref{deltaN-III}), we obtain, to linear order,
\be
\zeta_{r,{\rm TI}} \simeq \psi_{\rm QCD}-\psi_{\rm EW}  \simeq  -\frac{b}{b +1/2} \frac{R}{4} \times  \left. \frac{\delta V_{A0}}{\bar{V}_{A0}} \right|_{\rm exit} ~, \label{deltaN-III'}
\ee
as the dominant contribution to the uniform temperature curvature perturbation after the QCD transition during the thermal inflation $\zeta_{r,{\rm TI}}$.
Compared with Eq.~(\ref{scalar amplitude}), this is smaller by a factor $b/(b+1/2)$ which goes to unity in the instantaneous energy transfer limit $b \to \infty$.

%Let $\zeta_{r,{\rm TI}}$ denote the uniform temperature curvature perturbation after the QCD transition during the thermal inflation.
Since the thermal inflation ends on the uniform temperature slice characterized by $T=T_{\rm end}^{}$,
scale factor's perturbation on this ``end'' slice is given as
\be
\psi_{\rm end} = - \zeta_{r,{\rm TI}}  ~.
\ee
From the assumption, we have
\be
V_{0} = \rho_{\phi}|_{\rm end} ~.
\ee
Therefore, $\rho_{\phi} = \bar{\rho}_{\phi} \exp (-3(\zeta_{\phi}+\psi))$ is uniform on this slice, and thus,
\be
\zeta_{\phi} = - \psi_{\rm end} 
\ee
to conserve until $\phi$'s decay.
On the ``decay'' slice, the Hubble rate is uniform.
In other words,
\be
\delta \rho |_{\rm decay} = 0 ~.
\ee
We further assume that the energy density is dominated by the coherent oscillation before this slice.
This is actually the case with the sufficiently long thermal inflation discussed in Section \ref{thermal inflation}.
Then, $\Omega_{\phi} \simeq 1$ and all the other contributions are negligible in the curvature perturbation Eq. (\ref{linear-curvature-perturbation-perfect-fluids}). 
Therefore, we get
\be
\psi_{\rm decay} \simeq -\zeta_{\phi} ~.
\ee
After the decay slice, the radiation as the decay product is dominant component.
Therefore,
\be
\zeta_{r,{\rm decay}} \simeq - \psi_{\rm decay} \simeq \zeta_{\phi} = - \psi_{\rm end} = \zeta_{r,{\rm TI}} ~,
\ee
that is to say, if the thermal inflation lasts long enough, the curvature perturbation associated with the radiation existing during the thermal inflation is copied to the one after the reheating.

Provided that the temperature of this decay product is higher than the EW one, we can follow the discussion in \ref{app:gradual} but now with $\zeta_{r,{\rm decay}}$ replacing $\zeta_{\rm inf}$.
Neglecting the tiny value of $R := \bar{x}_{\rm QCD}$ in the standard thermal history realized after the thermal inflation, we finally obtain the observable one as $\zeta_{r} \simeq \zeta_{r,{\rm decay}}\simeq \zeta_{r,{\rm TI}}$.

\

%%%%%%%%%%%%%%%%%%%%%%%%%%%%%%%%%%%%%%%%%%
%%%%%%%%%%%%%%%%%%%%%%%%%%%%%%%%%%%%%%%%%%
\section{Isocurvature non-Gaussianity} \label{app:isocurvatureNG}
In order to evaluate the non-Gaussianity from the isocurvature perturbations as well as its power spectrum,
axion's evolution equation is integrated from the QCD slice to the end slice, resulting in Eq.~(\ref{theta_end})
 for a sufficiently large e-folding number $\Delta_{\rm end}$. 
With $\theta_{\rm end} \ll 1$, the evolution after the end of the thermal inflation is almost linear, and thus, we have $[\delta \theta /\bar{\theta} ]_{\rm late~time} \simeq [\delta \theta /\bar{\theta} ]_{\rm end}$ in Eq.~(\ref{late time S_A}) and 
\aln{S_{A} \simeq 2 \frac{\delta \theta_{\rm end}}{\bar{\theta}_{\rm end}} - \left( \frac{\delta \theta_{\rm end}}{\bar{\theta}_{\rm end}} \right)^{2} + \cdots ~,}
where the second order term is also retained since, in the following, we compute the nonlinear parameters defined by Eq.~(\ref{fNL}).  
Let us remember that the non-Gaussianity of 
  the purely adiabatic component Eq.~(\ref{f(yne0)}) is enhanced by $1/R$, as in Eq.(\ref{f(y=0)}). 
This is also the case in other components.
Here, we focus on the largest power of $1/R$ for each of $\Phi^{\rm a, i}_{\theta}$ and $\Phi^{\rm a, i}_{\theta \theta}$.

The isocurvature contribution to the gravitational potential defined in Eq.~(\ref{Phi^a Phi^i}) is given as 
\aln{\Phi^{\rm i}  \simeq \Phi^{\rm i}_{\theta}  \delta \theta_{\rm ini}+ \frac{\Phi^{\rm i}_{\theta \theta} }{2} \delta \theta_{\rm ini}^{2} + {\cal O}(\delta \theta_{\rm ini}^{3}) ~, \label{Phi_i}}
\aln{\Phi^{\rm i}_{\theta} = \overline{\frac{\delta \theta_{\rm QCD}}{\delta \theta_{\rm ini}} } \Phi^{\rm i}_{\theta,{\rm QCD}} ~, ~~~~\Phi^{\rm i}_{\theta \theta} = \left( \overline{\frac{\delta \theta_{\rm QCD}}{\delta \theta_{\rm ini}} } \right)^{2} \Phi^{\rm i}_{\theta \theta,{\rm QCD}} + \overline{\frac{\delta^{2} \theta_{\rm QCD}}{\delta \theta_{\rm ini}^{2}}} \Phi^{\rm i}_{\theta,{\rm QCD}} ~, \nn }
where
\aln{ \Phi^{\rm i}_{\theta,{\rm QCD}}  := \frac{r_{A} r_{c}}{5} \overline{\frac{2}{\theta_{\rm end}} \frac{\delta \theta_{\rm end}}{\delta \theta_{\rm QCD}}}= \frac{r_{A} r_{c}}{5}  \frac{2 {\cal Z}(\bar{\theta}_{\rm QCD})}{\bar{\theta}_{\rm QCD} }   \simeq \frac{r_{A} r_{c}}{5}  \frac{2}{\sin (\bar{\theta}_{\rm QCD}) }  ~,  \nn }
\aln{ \Phi^{\rm i}_{\theta \theta,{\rm QCD}} := \frac{r_{A} r_{c}}{5} \overline{ \frac{\delta}{\delta \theta_{\rm QCD}} \left(\frac{2}{\theta_{\rm end}} \frac{\delta \theta_{\rm end}}{\delta \theta_{\rm QCD}}\right) } \simeq  \frac{r_{A} r_{c}}{5} \frac{-2 \cos (\bar{\theta}_{\rm QCD}) }{\sin^{2} (\bar{\theta}_{\rm QCD}) }    ~.  \nn }
Note that, in evaluating the approximate equalities, ${\cal O}(R)$ term are omitted. 
See the discussion around Eq.~(\ref{''delta N''}). 
Similarly, the adiabatic contribution to the gravitational potential is given as
\be
\Phi^{\rm a}  &\simeq & \Phi^{\rm a}_{\theta}  \delta \theta_{\rm ini} + \frac{\Phi^{\rm a}_{\theta \theta} }{2} \delta \theta_{\rm ini}^{2} + {\cal O}(\delta \theta_{\rm ini}^{3}) ~, \label{Phi_a}
\ee
\aln{
\Phi^{\rm a}_{\theta} = -\frac{3}{5} \frac{\delta \zeta_{r{\rm G}}}{\delta \theta_{\rm ini}} = \overline{\frac{\delta \theta_{\rm QCD}}{\delta \theta_{\rm ini}} } \Phi^{\rm a}_{\theta,{\rm QCD}}  ~, \nn
}
\aln{
\Phi^{\rm a}_{\theta \theta} = 2 \left(\Phi^{\rm a}_{\theta} \right)^{2} f_{\rm NL}(\bar{\theta}_{\rm QCD},y) = \left( \overline{\frac{\delta \theta_{\rm QCD}}{\delta \theta_{\rm ini}} } \right)^{2} \Phi^{\rm a}_{\theta \theta,{\rm QCD}} + \overline{\frac{\delta^{2} \theta_{\rm QCD}}{\delta \theta_{\rm ini}^{2}}} \Phi^{\rm a}_{\theta,{\rm QCD}} ~,  \nn
}
where
\aln{
\Phi^{\rm a}_{\theta,{\rm QCD}} := -\frac{3}{5} \frac{R {\cal X}(\bar{\theta}_{\rm QCD})}{ 2 \bar{\theta}_{\rm QCD} }  = - \frac{3}{5} \frac{R}{4} \frac{\sin (\bar{\theta}_{\rm QCD}) }{1- \cos (\bar{\theta}_{\rm QCD})} ~ , \nn
}
\aln{
\Phi^{\rm a}_{\theta \theta,{\rm QCD}} := 2 \left(\Phi^{\rm a}_{\theta,{\rm QCD}} \right)^{2} f_{\rm NL}(\bar{\theta}_{\rm QCD},0) \simeq - \frac{3 R}{20} \frac{\cos (\bar{\theta}_{\rm QCD}) }{1- \cos (\bar{\theta}_{\rm QCD})} ~.
}
Comparing Eq.~(\ref{Phi_i}) and Eq.~(\ref{Phi_a}), one finds 
\aln{
\Phi^{\rm i}_{\theta} = - {\cal W} \times \Phi^{\rm a}_{\theta} ~,\label{def of W}
}
\aln{ {\cal W}  := \frac{4 r_{A} r_{c}}{3 R} \frac{{\cal Z}(\bar{\theta}_{\rm QCD})}{{\cal X}(\bar{\theta}_{\rm QCD})}  \simeq  \frac{8 r_{A} r_{c}}{3 R } \frac{1}{1+ \cos (\bar{\theta}_{\rm QCD})}~.\label{explicit W}
}
This factor is nothing but the ratio of Eq.~(\ref{condition 1}) to Eq.~(\ref{condition 2}), multiplied by the numerical factor $r_{c}/3 \simeq 0.28$ coming from the definition Eq.~(\ref{Phi^a Phi^i}). 
On the other hand, 
\aln{
\Phi^{\rm i}_{\theta\theta,{\rm QCD}} = + {\cal W}\times \Phi^{\rm a}_{\theta\theta,{\rm QCD}} ~,
}
up to ${\cal O}(r_{A} R^{0})$ terms. 
Therefore, we get
\aln{
\Phi^{\rm i}_{\theta\theta} &\simeq + {\cal W} \times \Phi^{\rm a}_{\theta\theta} -2 {\cal W}\overline{\frac{\delta^{2} \theta_{\rm QCD}}{\delta \theta_{\rm ini}^{2}} \frac{\delta \theta_{\rm ini}}{\delta \theta_{\rm QCD}}} \times \Phi^{\rm a}_{\theta}   ~, \nn \\
&\simeq + {\cal W} \times \Phi^{\rm a}_{\theta\theta} -2 {\cal W} y \sin (\bar{\theta}_{\rm QCD}) \times \Phi^{\rm a}_{\theta}
}
On the second line, ${\cal O}(y^{2})$ terms are neglected.

From the constraint Eq.~(\ref{r and R}), we have
\aln{ {\cal W} < 9.2 \times 10^{-3} ~. \label{bound on W}
}
This means that, once the constraint on the purely adiabatic component Eq.~(\ref{f and g}) is met, then 
\aln{ f^{\rm a,ai}_{\rm NL} = - {\cal W} f^{\rm a,aa}_{\rm NL} ~, ~~~  f^{\rm a,ii}_{\rm NL} =  +{\cal W}^{2} f^{\rm a,aa}_{\rm NL} \label{relation of f}
}
turn out to be sufficiently small. 
Other components with $\Phi^{\rm i}_{\theta\theta}$ contain terms that are not written in terms of $f^{\rm a,aa}_{\rm NL}$:
\aln{ f^{\rm i,aa}_{\rm NL} = + {\cal W} f^{\rm a,aa}_{\rm NL} - \frac{y {\cal W}}{\Phi^{\rm a}_{\theta}} ~, ~~  f^{\rm i,ai}_{\rm NL} = - {\cal W}^{2} f^{\rm a,aa}_{\rm NL} + \frac{y {\cal W}^{2}}{\Phi^{\rm a}_{\theta}} ~, ~~  f^{\rm i,ii}_{\rm NL} =  {\cal W}^{3} f^{\rm a,aa}_{\rm NL} - \frac{y {\cal W}^{3}}{\Phi^{\rm a}_{\theta}}  }
for $\bar{\theta}_{\rm QCD} \simeq \pi/2$, and thus, new constraints on $y$ may in principle appear. 
However, for the moment, the Planck experiment gives $f^{\rm i,aa}_{\rm NL} = 96 \pm 52$ at 68\%CL and the weaker ones on the other two \cite{Akrami:2019izv}.  
With $\Phi^{\rm a}_{\theta} \simeq -3 R/20$ in the denominator, $y<1$ turns out to be  sufficient to meet the observational constraints for the isocurvature non-Gaussianity.

%%%%%%%%%%%%%%%%%%%%%%%%%%
\bibliography{Bibliography}
    \bibliographystyle{unsrt}
    
\end{document}